\documentclass[aps,prr,twocolumn,titlepage,nofootinbib]{revtex4}
\usepackage{graphicx}
\usepackage{color}
\usepackage{dcolumn}
\usepackage{latexsym}

\usepackage[normalem]{ulem}
\usepackage{hyperref,amssymb}
\usepackage{url}
\usepackage{color,soul}
\usepackage{graphicx}
\usepackage{verbatim}
\usepackage{multirow}
\usepackage{amsmath}
\newcommand{\beq}{\begin{eqnarray}}
\newcommand{\eeq}{\end{eqnarray}}
\usepackage{mathrsfs}
\usepackage{float,soul}
\usepackage[dvipsnames]{xcolor}
\usepackage{mathtools}
\usepackage{slashed}
\usepackage{physics}
\usepackage{graphicx}
\usepackage{epstopdf}
\usepackage{subfigure}
\usepackage[font=small]{caption}
\usepackage{hyperref}
\usepackage{bbold}
\usepackage{wasysym}
\usepackage{feynmp}
\usepackage{hyperref}
\hypersetup{colorlinks}
\usepackage{ragged2e}

\usepackage[latin1]{inputenc}

\begin{document}

\title{Braids and Higher-Order Exceptional Points from the Interplay Between Lossy Defects and Topological Boundary States}
\author{Zi-Jian Li$^{1,\dag}$}
\author{Gabriel Cardoso$^{2,\dag}$}
\author{Emil J. Bergholtz$^{3}$}
\author{Qing-Dong Jiang$^{1,2,4}$}
\email{qingdong.jiang@sjtu.edu.cn}
\affiliation{{}\\$^1$ School of Physics and Astronomy, Shanghai Jiao Tong University, Shanghai 200240, China\\
$^2$ Tsung-Dao Lee Institute, Shanghai Jiao Tong University, Shanghai 200240, China\\
$^3$ Department of Physics, Stockholm University, AlbaNova University Center, 106 91 Stockholm, Sweden\\
$^4$ Shanghai Branch, Hefei National Laboratory, Shanghai 201315, China\\
$^\dag$ These authors contributed equally to this work
}

\begin{abstract}
We show that the perturbation of the Su-Schrieffer-Heeger chain by a localized lossy defect leads to higher-order exceptional points (HOEP). Depending on the location of the defect, third- and fourth- order exceptional points (EP3 \& EP4) appear in the space of Hamiltonian parameters. On the one hand, they arise due to the non-Abelian braiding properties of exceptional lines (EL) in parameter space. Namely, the HOEPs lie at intersections of mutually non-commuting ELs. On the other hand, we show that such special intersections happen due to the fact that the delocalization of edge states, induced by the non-Hermitian defect, hybridizes them with defect states. These can then coalesce together into an EP3. When the defect lies at the midpoint of the chain, a special symmetry of the full spectrum can lead to an EP4. In this way, our model illustrates the emergence of interesting non-Abelian topological properties in the multiband structure of non-Hermitian perturbations of topological phases. 

\end{abstract}
\maketitle

\section{Introduction}

Lattice defects have long been known to give important information about crystalline phases \cite{andreev1969quantum,bollmann2012crystal}. When the lattice has a topological nature, the domain walls which appear at defects trap localized zero modes \cite{hasan2010colloquium}. Non-Hermitian generalizations arise when the lattice and/or the defects are non-Hermitian, which vastly extends the number of emergent phenomena. For example, non-Hermitian systems are known to display the famously anomalous bulk-boundary correspondence and the non-Hermitian skin effects \cite{Yao2018,Lee2016,Xiong2018,Kunst2018,xiao2020non,helbig2020generalized,ghatak2020observation,Borgnia,scalefree21,Okuma2023}, and adding a single defect may lead to topological defect states \cite{Schomerus:13,poli2015selective}, the breakdown of skin effects \cite{jana2023emerging} and the formation of anomalous skin effects \cite{scalefree23,scalefree232,molignini2023}. 
In Hermitian lattices, adding non-Hermitian defects also leads to unconventional phenomena, such as exceptional points (EP)\cite{jin2017schrieffer,tzortzakakis2022topological,burke2020non}. In this work we find that, when the Hermitian lattice is topological, its perturbation by a single non-Hermitian defect can lead to higher-order exceptional points (HOEP). We analyse the multi-band spectrum and the resulting non-Abelian braiding topology of EPs to clarify its connection to the topological edge states.

\begin{figure}
\centering
\includegraphics[width=1\linewidth]{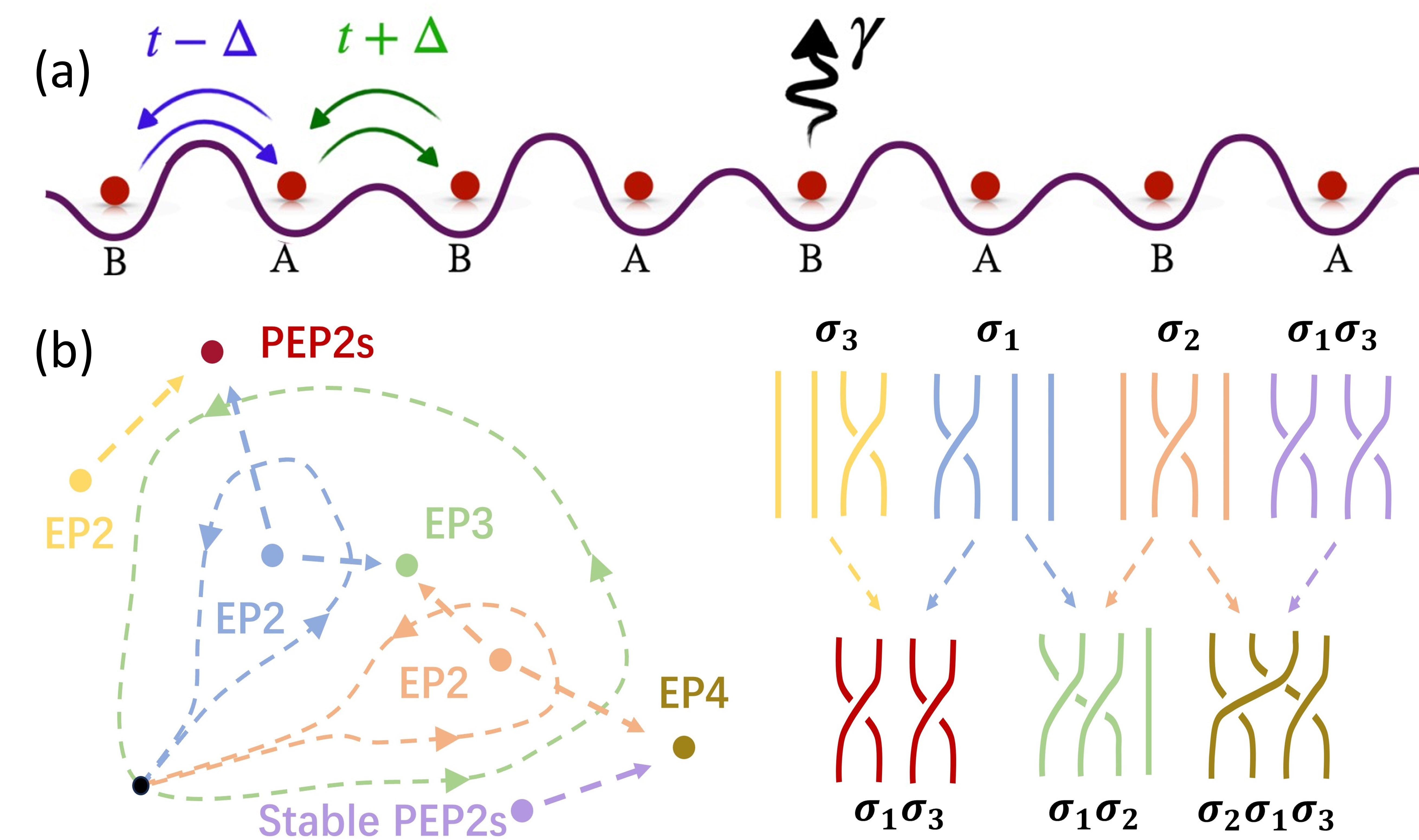}

\caption{\justifying {(a) Schematic diagram of the Su-Schrieffer-Heeger chain with a lossy defect. (b) Formation of PEP2s, an EP3, and an EP4 from the merging of EPs. From the braiding of eigenvalues (right) as one traces a loop around an EP (left), one sees that these three different cases arise from the different numbers of common states between the braidings on each merging EP.}}
\label{fig1}
\end{figure}

EPs are unconventional spectral singularities where the Hamiltonian becomes defective due to the coalescence of eigenstates \cite{heiss2012physics,heiss2004exceptional,berry2004,ashida2020non,Bergholtz2021}. They have been observed in various experimental platforms such as electronic circuits \cite{stehmann2004observation}, microwave cavities \cite{dembowski2001experimental}, acoustics \cite{shi2016accessing}, photonics \cite{ozdemir2019parity,miri2019exceptional}, and nitrogen-vacancy systems \cite{liu2021dynamically,wu2024third}, and were shown to enhance superconductivity \cite{arouca2023topological}, speed up entanglement generation \cite{li2023speeding} and enhance quantum heat engines \cite{bu2023enhancement}. In multi-band systems, higher-order exceptional points (HOEP) can appear, where $N>2$ states simultaneously coalesce, also known as $N$-th order exceptional points (EP$N$). Symmetry and topology play a major role in classifying and protecting EPs \cite{ding2022non,patil2022measuring,delplace2021symmetry, mandal2021symmetry,carlstromKnottedNonHermitianMetals2019,zhangTidalSurfaceStates2021, hu2021knots, zhong2018winding,hu2022knot,rui2019topology,luitz2019exceptional,zhou2018observation,guo2023exceptional,konig2023braid,wojcik2022eigenvalue,lin2019symmetry,gonzalez2017topological,ryu2012analysis,kozii2017non,Hu2023,Yang2023,budichSymmetryprotectedNodalPhases2019a, kawabata2019classification,kimuraChiralsymmetryProtectedExceptional2019, okugawaTopologicalExceptionalSurfaces2019, szameitMathcalMathcalSymmetry2011, yoshidaSymmetryprotectedExceptionalRings2019,PhysRevB.107.144304, zhouExceptionalSurfacesPTsymmetric2019,stalhammarClassificationExceptionalNodal2021,crippaFourthorderExceptionalPoints2021,wangExperimentalSimulationSymmetryprotected2023,halder2022properties} and nodal points \cite{bouhon2020geometric,bouhon2020non}, with the associated topological invariant being the eigenvalue braiding on an encircling loop in the space of Hamiltonian parameters. It corresponds to an element of the non-Abelian braid group \cite{wojcik2020homotopy, li2021homotopical, delplace2021symmetry,kawabata2019classification}, and in particular the formation of HOEPs can be accounted for in terms of the merging of non-commuting EP2s. We show that this structure naturally appears in the spectrum of the Su-Schrieffer-Heeger (SSH) chain perturbed by a localized lossy defect, and a rich landscape of exceptional degeneracies appears in the 3D parameter space formed by the (real) SSH gap $\Delta$ and the (complex) defect strength $\gamma$ (figure \ref{fig1}(a)). If the defect is located at: the edge site, a generic bulk site, or the mid-point of the chain, then we find that: paired second-order exceptional points (PEP2) where EP2s connecting different states appear at the same point, an EP3, or an EP4, respectively, appear. We classify them in terms of the non-Abelian merging of EP2s with different braid invariants, as schematically represented in figure \ref{fig1}(b). One finds that the HOEPs only appear in the topological sector of parameter space $\Delta>0$. We explain this fact by studying the effect of the defect on the topological edge states. We find that, in the multi-band context, the previously noted delocalization of edge states can lead to their coalescence with defect states, which in parameter space translates into the merging of non-commuting EPs into a HOEP. This reveals a new mechanism by which non-Abelian non-Hermitian topology can arise from localized perturbations of topological phases.

The paper is structured as follows. In section \ref{sec:model} we introduce the model and the spectral degeneracies. Section \ref{sec:ELs} reviews the method of resultants and proceeds to apply it to study the exceptional lines (EL) in the parameter space of our model. In section \ref{NAB}, we study the non-Abelian structure of the ELs. For completeness, we first review how it arises from the braid invariant construction, which we than apply to the HOEPs in the parameter space of our model. In section \ref{roleofedge}, we trace the appearance of the HOEPs back to the interplay between edge states and the lossy defect. Finally, we discuss the main conclusions, generalizations and future directions in section \ref{sec:discussion}. Appendices are included for extra computational details.

\section{The Model}\label{sec:model}
We consider a perturbation of the SSH chain by a single lossy defect at odd site $2s-1$,
\begin{equation}
    H = H_0 - i\gamma |2s-1 \rangle \langle 2s-1|,\label{eq:defectH}
\end{equation}
where
\begin{equation}
    H_0 = -\sum_{j=1}^{2N-1} (t +(-1)^j\Delta)|j\rangle \langle j+1| +  \text{h.c.}\label{eq:SSH}
\end{equation}
is the SSH Hamiltonian on a chain of length $2N$, and we investigate the non-dimerized regime $|\Delta|<1$. This model can be realized in dielectric resonator chains and in photonic lattices, and in both experimental setups the Hamiltonian spectrum can be directly measured (see, for example, \cite{poli2015selective} and \cite{Schomerus:13}, respectively). Note that one can recover the case of a defect at an even site by reversing the chain. Also, since the spectrum is symmetric with respect of the sign of $\gamma$, we focus on the dissipative regime $\text{Re}(\gamma)>0$. In our notations, the hopping strength is normalized to $t=1$. The topologically nontrivial phase corresponds to $\Delta>0$, for which the spectrum of (\ref{eq:SSH}) contains edge states protected by chiral symmetry \cite{SSHreview}. As we will see, the structure of non-Hermitian degeneracies is also different in this phase.

\subsection{Spectral Properties}
\begin{figure}
\centering
\includegraphics[width=1\linewidth]{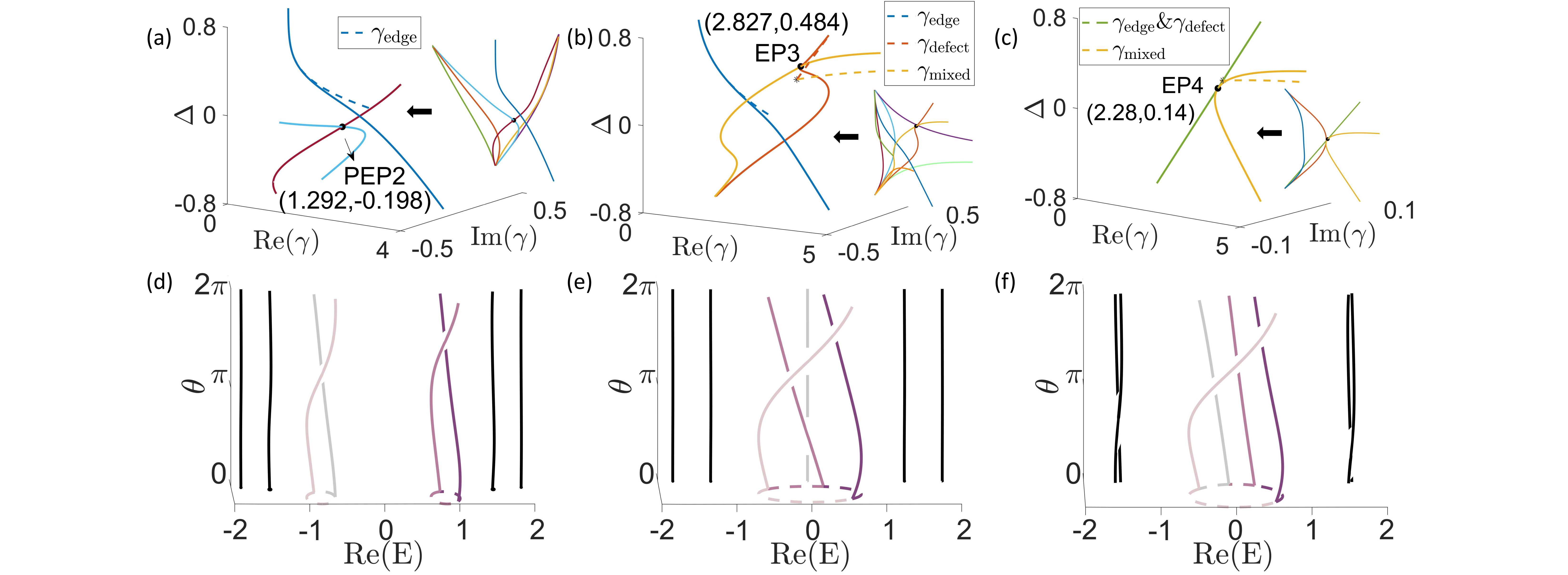}
\caption{\justifying{ Complex spectra of the model with $\Delta = 0.3, s=3, 2N=16
$. The colored lines are the levels displaying exceptional degeneracies. The inset shows the characteristic square-root branch point for the coalescing levels at small $\gamma$. }}
\label{spectrum}
\end{figure}

The first indication of EPs appears in the degeneracies of the eigenvalue spectrum. We numerically evaluate it as a function of the defect strength $\gamma$ (with fixed $\Delta$) through exact diagonalization. In figure \ref{spectrum}, we notice three degeneracies, located at different values of $\gamma$. These correspond to square-root branch points of the eigenvalues (see the inset), at which two eigenvectors coalesce. This type of degeneracy of non-Hermitian Hamiltonians is known in various contexts \cite{dykhne1960quantum,hwang1977adiabatic,wang2022adiabaticity,cardoso2023landau} and corresponds to an EP2.
If two or more of these degeneracy points can be made to coincide, for example by tuning also the $\Delta$ parameter, then higher-order degeneracy points appear, where more eigenvalues simultaneously coalesce. We also note that, where this happens, the coalescing eigenvalues themselves are special. Namely, two of the eigenvalues involved are in-gap states, while the others come from different bands. This is the first indication of the role of topological edge states in leading to higher-order EPs, as we will explore in detail in the following sections.

As we now discuss, one can more effectively scan the parameter space for spectral degeneracies, effectively bypassing the evaluation of the full spectrum.

\section{Line degeneracies and higher-order degeneracies}\label{sec:ELs}
\subsection{Resultants calculation}
The useful mathematical construct is that of polynomial resultants. Given two single-variable polynomials $p(z) = a_n z^n + \cdots + a_0$ and $q(z) = b_m z^m + \cdots + b_0$, their resultant is defined as
\begin{equation}
    {\rm Res}_z (p,q) = a_n^m \prod_{i=1}^n q(z_i),
\end{equation}
where $\{{z_i}\}$ is the full set of roots of the polynomial $p(z)$. The definition is such that, if $p$ and $q$ have a common root, then their resultant vanishes.
A convenient result is that the resultant can be evaluated through the determinant of the Sylvester matrix $\mathcal S(p,q)$ \cite{janson2007resultant,woody2016polynomial},
\begin{equation}
    {\rm Res}_z (p,q) = {\rm det}\ {\mathcal S} (p,q), \label{res_syl}
\end{equation}
where $\mathcal S(p,q)$ is the $(m+n)\times (m+n)$ matrix
\begin{equation}
\mathcal S(p,q) = \left(
    \begin{matrix}
        a_n & a_{n-1} & a_{n-2} & \ldots & 0 & 0 & 0 \\
        0 & a_{n} & a_{n-1} & \ldots & 0 & 0 & 0 \\
        \vdots & \vdots & \vdots & & \vdots& \vdots& \vdots \\
        0 & 0 & 0 & \ldots & a_1 & a_0 & 0 \\
        0 & 0 & 0 & \ldots & a_2 & a_1 & a_0 \\
        
        b_m & b_{m-1} & b_{m-2} & \ldots & 0 & 0 & 0 \\
        0 & b_{m} & b_{m-1} & \ldots & 0 & 0 & 0 \\
        \vdots & \vdots & \vdots & & \vdots& \vdots& \vdots \\
        0 & 0 & 0 & \ldots & b_1 & b_0 & 0 \\
        0 & 0 & 0 & \ldots & b_2 & b_1 & b_0 \\
    \end{matrix}
    \right). \label{syl}
\end{equation}

Recall that diagonalization of the Hamiltonian can be stated as the algebraic problem of finding the roots of the characteristic polynomial
\begin{equation}
    p(\lambda;\mathbf R) \equiv \det (H(\mathbf R) - \lambda I) \label{eq:characp}
\end{equation}
in the variable $\lambda$, where the coefficients of the polynomial depend on the Hamiltonian parameters $\mathbf{R}$. Therefore, degeneracies (if they exist) correspond to repeated roots of the polynomial $p(\lambda;\mathbf{R})$. In particular, a $k$-th order degeneracy is given by a root $\lambda^*$ of multiplicity $k$, which also implies the vanishing of the derivatives of the characteristic polynomial up to order $k-1$,
\begin{equation}
    \begin{aligned}
        & p(\lambda^*) = p'(\lambda^*) = \cdots = p^{(k-1)}(\lambda^*) = 0, \\
        & p^{(k)}(\lambda^*) \neq 0.
    \end{aligned}
\end{equation}
The last inequality guarantees that $\lambda^*$ is not a root of multiplicity $k+1$. Finally, these equations mean that $\lambda^*$ is a common root between $p$ and its derivatives, which translates to a condition on the resultants,
\begin{equation}
    \begin{aligned}
        & {\rm Res}_{\lambda} \left(p(\lambda),p^{(l)}(\lambda)\right) = 0, \quad l=1,\cdots,k-1, \\
        & {\rm Res}_{\lambda} \left(p(\lambda),p^{(k)}(\lambda)\right) \neq 0.
    \end{aligned}
    \label{res}
\end{equation}
These give $k-1$ polynomial equations on the parameters $\mathbf{R}$. Since the resultants are complex expressions, one in general needs to tune $2(k-1)$ real parameters to locate a $k$-th order degeneracy \cite{mandal2021symmetry,delplace2021symmetry,sayyad2022realizing}.

\begin{figure}
\centering
\includegraphics[width=1\linewidth]{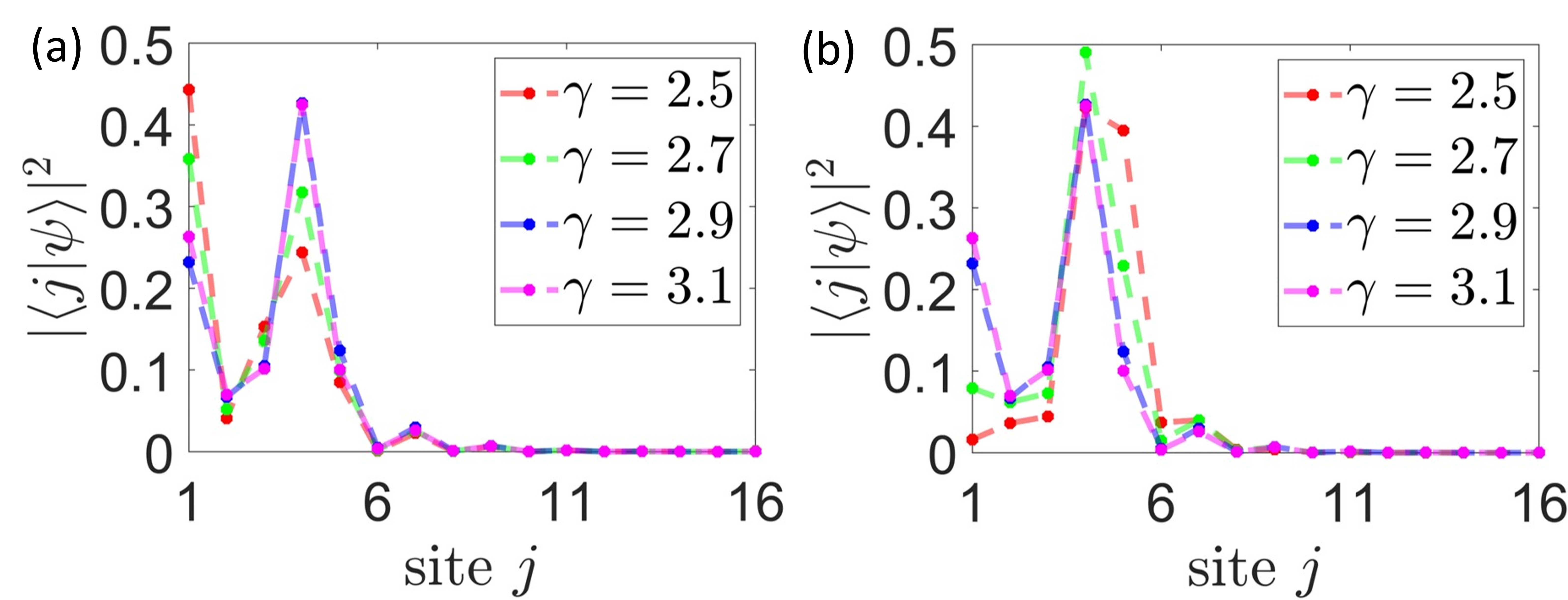}
\caption{\justifying Line degeneracies in 3D parameter space $(\Delta,{\rm Re}(\gamma),{\rm Im}(\gamma))$ for the SSH chain of different lengths with a lossy defect located at different sites. The black dots are the intersecting points of two exceptional lines in all figures. (a)\&(b) The defect is located at the boundary $s=1$. The lengths of the chain are $2N=8,12$ respectively. (c)\&(d) The defect is located at a generic bulk site. For figure (c), the length and the defect location are $2N=8,s=2$; for figure (d), the length and the defect location are $2N=10,s=3$. (e)\&(f) The defect is located at the ``midpoint" of the chain $s=\frac{N}{2}+1$, where $N$ is even. The lengths of the chain are $2N=8,12$ respectively.}
\label{ELs}
\end{figure}

\subsection{Exceptional lines}
Interestingly, we have found a simple analytical expression for the characteristic polynomial of the Hamiltonian (\ref{eq:defectH}) (see Appendix \ref{charac} for detailed derivation):
\begin{equation}
\begin{aligned}
   & p(\lambda;\Delta, \gamma) = (1-\Delta^2)^{N}\times \\
   & \left[(U_N + r U_{N-1}) + \frac{i\gamma \lambda}{1-\Delta^2} (U_{s-1} + r U_{s-2})U_{N-s}\right],
\end{aligned}
\label{eq:characp_SSH}
\end{equation}
where $r = \frac{1+\Delta}{1-\Delta}$, $U_{n} = U_{n}(Q)$ is the $n$-th Chebyshev polynomial of the second kind \cite{mason2002chebyshev}, and $Q = \frac{\lambda^2 - 2(1+\Delta^2)}{2(1-\Delta^2)}$.

We solve for degeneracies by computing the corresponding resultants in the 3D parameter space $(\Delta,{\rm Re}(\gamma),{\rm Im}(\gamma))$ (for more details, see Appendix \ref{ld}). For simplicity, we focus on the two-fold degeneracy, which is given by the vanishing of a single resultant. This leads to two real equations, so that the degenerate points (for $\Delta$ fixed) extend into lines, as plotted in figure \ref{ELs}. We find that there are three qualitatively different types of degenerate line structures and intersections, corresponding to the three different defect locations (in the boundary, in the bulk, or in the midpoint of the chain). Up to this distinction, the relevant features are insensitive to the exact value of the chain length $2N$ and the defect location parameter $s$.

Although most of the degenerate lines are complicated expressions found by solving polynomials of large degree (which we do numerically), we find a symmetric configuration in which they are surprisingly simple. It happens when the defect is placed at the midpoint of the chain. If the total number of sublattice sites $N$ is even, $N=2m$, this corresponds to $s=m+1$. In this case, we show in Appendix \ref{paireddeg} that when $\gamma=2(1+\Delta)$ the characteristic polynomial becomes a perfect square
\begin{equation}
    p = \left(U_m + \frac{i\lambda+ 1+\Delta}{1-\Delta}U_{m-1}\right)^2.
    \label{eq:charac_evenN}
\end{equation}
This shows that the whole spectrum becomes degenerate in pairs at $\gamma = 2(1+\Delta)$. In fact, we will see that the eigenstates themselves coalesce in pairs, forming a paired exceptional point (PEP2). It is stable and extends into a line in 3D, represented as the green line in figures \ref{ELs}(e) and \ref{ELs}(f). We will also show that this special degeneracy accounts for the formation of an EP4 in the next section.

For $N=2m-1$ odd, and for $s=m$, one similarly has that on the line $\gamma = 2(1-\Delta)$ the characteristic polynomial is a perfect square
\begin{equation}
    p = r\left(\frac{\lambda+i(1-\Delta)}{1+\Delta}U_{m-1} + i U_{m-2}\right)^2,
    \label{eq:charac_oddN}
\end{equation}
and a PEP2 appears at $\gamma = 2(1-\Delta)$, which is exactly the dark blue line in \ref{ELs}(d). A similar case of PEP2 was reported in \cite{burke2020non}, which can be seen as a special case of the SSH chain when $\Delta=0$. We point out that the different expressions $\gamma=2(1\pm\Delta)$ for the PEP2 line and, as we will see, the appearance or not of the EP4, depend on the parity of the sublattice length $N$. This reveals a new instance of the even-odd effects characteristic of non-Hermitian systems \cite{joglekar2010robust, burke2020non}.

In principle, higher-order degeneracies can appear at the intersections of degeneracy lines, which are outlined in figure \ref{ELs}. However, just as one cannot determine just from the eigenvalue degeneracies whether these are exceptional lines (ie. whether the eigenvectors themselves coalesce), it is also not possible to assert that higher-order exceptional degeneracies (HOEPs) appear at the intersections. Instead of computing all the eigenvectors, we discuss a solution to this problem by decorating the exceptional lines with their braid invariants. These will properly qualify whether each degeneracy line is an exceptional line, and determine the nature of the HOEPs formed at their intersections.

\section{Non-Abelian braidings characterizing Degeneracies} \label{NAB}
\subsection{Braid group invariants}
Exceptional points (EPs) can be distinguished from normal degeneracies by the braiding of eigenvalues on a loop enclosing the degeneracy in parameter space (see figure \ref{fig1}(b) as an example) \cite{hu2022knot}. For a given EP of a generic $N$-band non-Hermitian Hamiltonian, this associates to each enclosing loop an element of the braid group $B_N$, whose conjugacy class is the topological invariant of the EP \cite{wojcik2020homotopy, li2021homotopical, delplace2021symmetry,kawabata2019classification}. More explicitly, the braid group element can be computed in the following way. First define a reference ordering of the eigenvalues (such as in terms of their real parts). Then, keeping track of the eigenvalues as the Hamiltonian parameters are adiabatically changed along the loop, take the crossing of the $i$-th eigenvalue over (respectively, under) the $i+1$-th eigenvalue to correspond to the $\sigma_i$ (resp., $\sigma_i^{-1}$) generator (c.f. \ref{fig1}(b)). The braid group is generated by the $\sigma_i$ with the rules \cite{fox1962braid}
\begin{equation}
\begin{aligned}
    & \sigma_i \sigma_{i+1} \sigma_i = \sigma_{i+1} \sigma_i \sigma_{i+1}\ (1\leq i \leq N-1),\\
    & \sigma_i \sigma_j = \sigma_j \sigma_i\ (|j-i|>1).
\end{aligned} \label{eq:braidgroupalgebra}
\end{equation}
Under continuous deformations of the loop which do not cross any additional EPs and the different choices of the reference ordering, the braiding only changes by a (type I) Markov move \cite{birman1974braids}, which corresponds to a conjugation of the braid group element 
\begin{equation}
    b_{\rm loop}\mapsto g b_{\rm loop} g^{-1},
\end{equation}
where $g\in B_N$. Thus the conjugacy classes $[b_{\rm loop}]$ are topological invariants of the EPs, and get multiplied when EPs merge \cite{guo2023exceptional}.

In three dimensions, the EP2s extend into ELs, and the braid invariants are computed from loops enclosing the ELs. They satisfy a non-Abelian conservation rule (NACR): invariants $[b]$ are constant as the loop moves along each isolated EL, and get multiplied when other ELs enter/leave the loop. In particular, if two ELs intersect forming a HOEP, then one can infer that there is a common state among the ones braiding around each EL. From equation (\ref{eq:braidgroupalgebra}), we see that the corresponding braid invariants do not commute. Conversely, if the braid invariants of the intersecting ELs do commute, then one can infer that the states braiding around each EL form two distinct pairs, and a HOEP is not formed. Instead, one has a PEP2. Both cases are schematically illustrated in figure \ref{fig1}(b), and we find that both happen in the defect SSH chain (\ref{eq:defectH}).

\subsection{Characterization of ELs and HOEPs}
\begin{figure}
\centering
\includegraphics[width=1\linewidth]{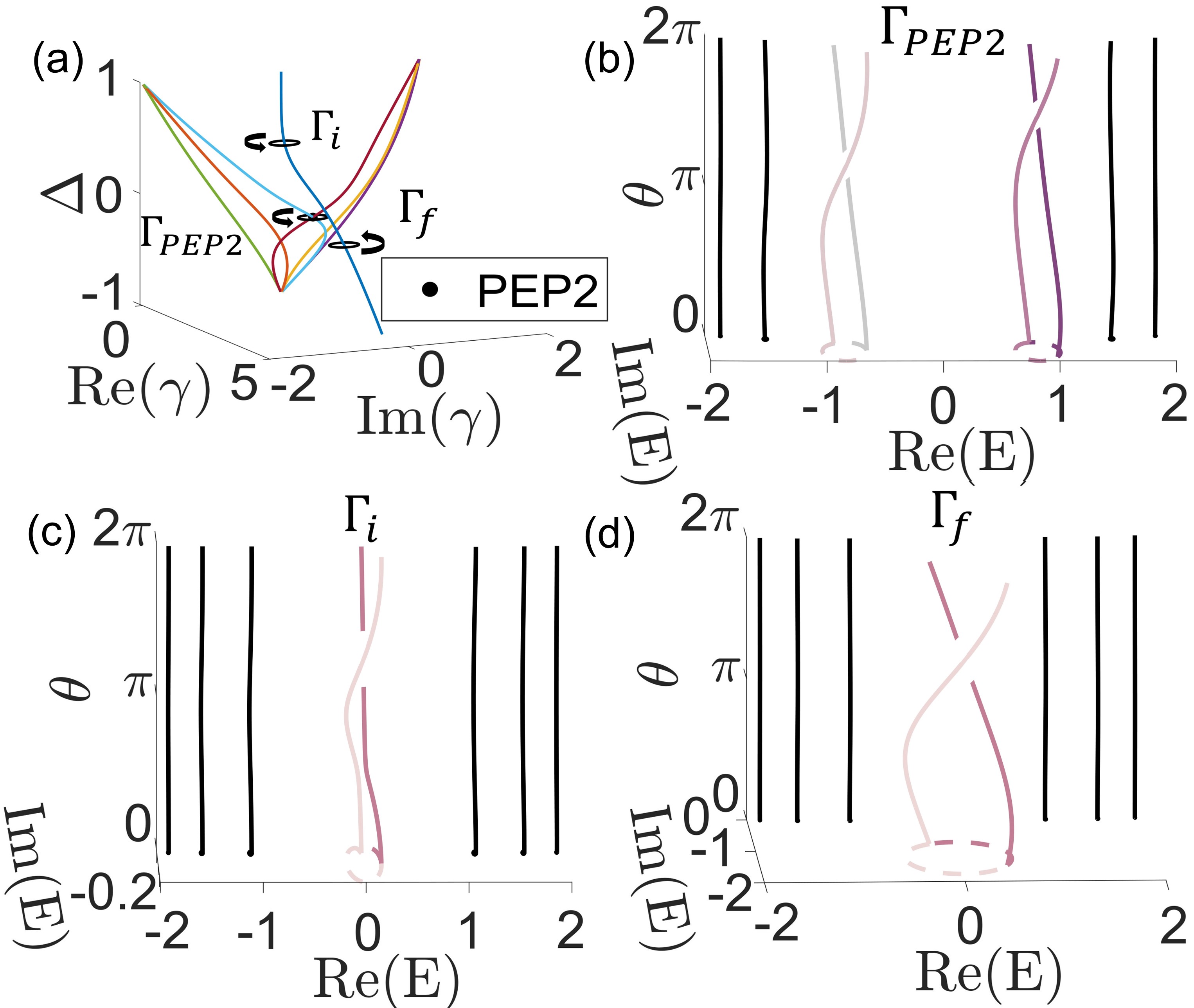}
\caption{\justifying Eigenvalue braiding and ELs for the SSH chain with a boundary defect ($s=1$) where $N=4,s=1$. (a) Closed paths encircling ELs and the PEP2 (black dot). (b) Eigenvalue braiding near the intersection (PEP2). The path is $\gamma = 1.25 + 0.2i e^{i\theta}$, $\Delta = -0.2$. (c) Eigenvalue braiding at $\Gamma_i$. The path is $\gamma = 0.13 + 0.2i e^{i\theta}$ and $\Delta = 0.36$. (d) Eigenvalue braiding at $\Gamma_f$. The path is $\gamma = 2.48 + 0.2i e^{i\theta}$ and $\Delta = -0.33$. For all figures, $\theta:0 \rightarrow 2\pi$.}
\label{boundary}
\end{figure}

As a simple illustration of the NACR and a verification of the exceptional degeneracy, we compute the eigenvalue braidings on two loops, $\Gamma_i$ and $\Gamma_f$, which both encircle the same degeneracy line, figure \ref{boundary}(a). The paths $\Gamma_i$ and $\Gamma_f$, which are loops at different fixed values of $\Delta$, can be continuously deformed into one another without intersecting other lines (ELs). According to the NACR, we expect that the braid invariant of the path remains unchanged. This is indeed the case, as shown in figures \ref{boundary}(c) and \ref{boundary}(d), where the braiding of eigenvalues on both the $\Gamma_i$ and $\Gamma_f$ loops is given by $\sigma_4$. In this way, one can assign to each EL a braid invariant. At the intersections of ELs, the invariants get multiplied, which can simply be shown by deforming a loop enclosing the intersection as a composition of loops enclosing each line in succession. The different combinations of invariants at the intersections characterize the different higher-order degeneracies which can appear. 

When the defect is located at the edge, $s=1$, we find that two ELs (magenta and light blue) intersect forming PEP2s, as shown in figure \ref{boundary}(a). Indeed, the braid invariants corresponding to the intersecting ELs commute, and any path enclosing both lines has braiding $[b_{\rm PEP2}] = [\sigma_3 \sigma_5]$ (figure \ref{boundary}(b)).

\begin{figure}
\centering
\includegraphics[width=1\linewidth]{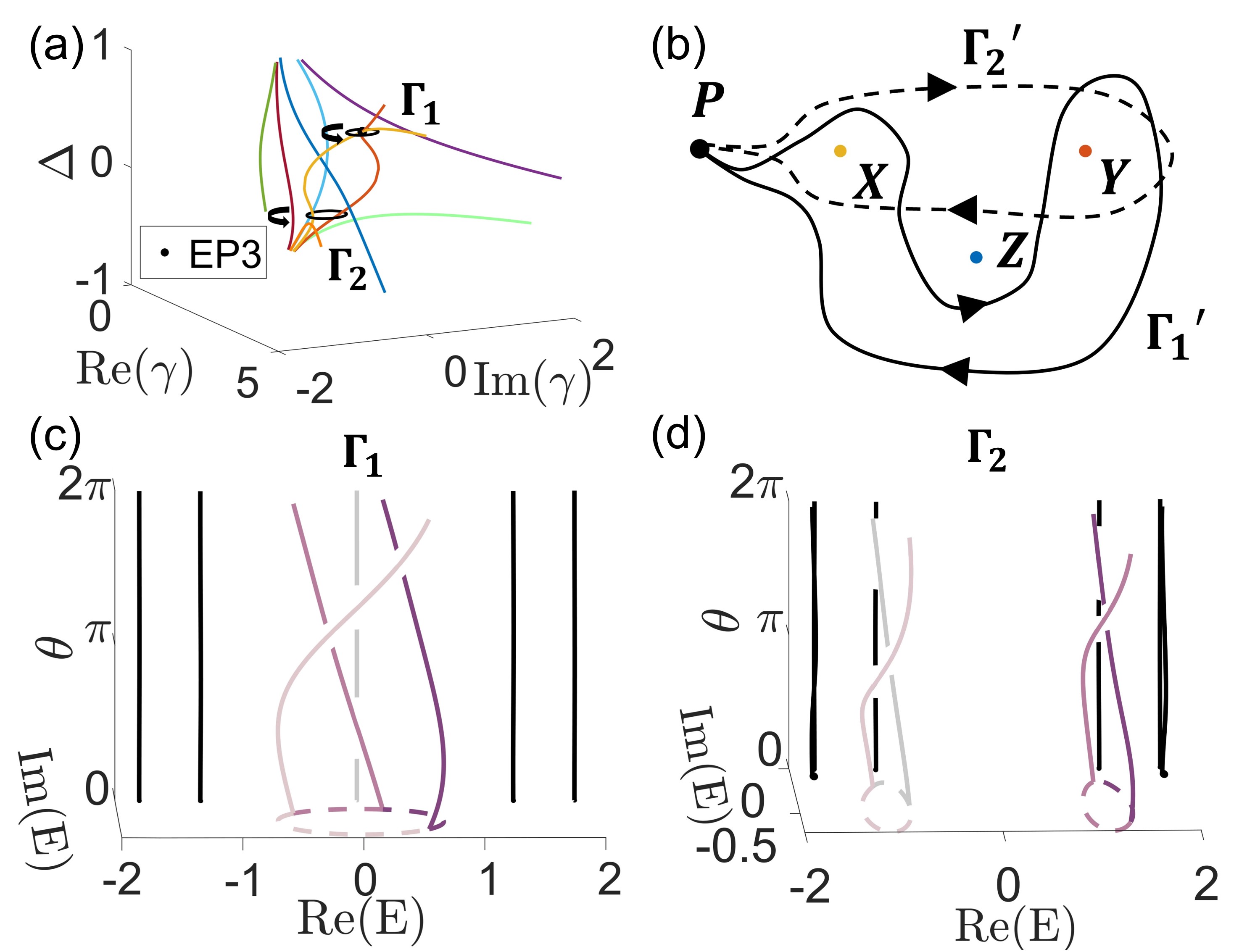}
\caption{\justifying Eigenvalue braiding and ELs for the SSH chain with a bulk defect ($1<s<\frac{N}{2}+1$), where $N=4,s=2$. (a) Closed paths encircling the PEP2 (brown dot) and the EP3 (black dot). (b) Topological distinction between paths $\Gamma_1'$ and $\Gamma_2'$ at an (arbitrary) intermediate $\Delta_m$ between the two intersections. (c) Eigenvalue braiding near the EP3. The path is $\gamma = 2.83 + 0.2i e^{i\theta}$ and $\Delta = 0.48$. (d) Eigenvalue braiding at the $\Delta<0$. The path is $\gamma = 1.62 + 0.25i e^{i\theta}$ and $\Delta = -0.36$. For all figures, $\theta:0 \rightarrow 2\pi$.}
\label{bulk}
\end{figure}

When the defect is located at a generic bulk site, an EP3 appears, located at the intersection of the orange and yellow lines in the $\Delta>0$ region, figure \ref{bulk}(a). The braiding on a loop enclosing the intersection, $\sigma_4^{-1}\sigma_3\sigma_4\sigma_5$, is shown in figure \ref{bulk}(c). We bring it to standard form by a Markov move,
\begin{equation}
    (\sigma_5\sigma_4)(\sigma_4^{-1}\sigma_3\sigma_4\sigma_5)(\sigma_5\sigma_4)^{-1}=\sigma_3\sigma_4, \label{move}
\end{equation}
where we used that $\sigma_5\sigma_4\sigma_5=\sigma_4\sigma_5\sigma_4$. Thus $[b_{\Gamma_1}]=[\sigma_3\sigma_4]$, with the product of non-commuting braids capturing the triple braiding of eigenvalues characteristic of an EP3 (see figure \ref{fig1}(b)).

The unique structure of ELs in figure \ref{bulk}(a) also demonstrates the non-local nature of the EPs. Although the two intersections at $\Delta<0$ and $\Delta>0$ regions are generated by the same ELs (the orange and yellow lines), they lead to different exceptional degeneracies. The difference is reflected in the braidings of eigenvalues, as shown in figures \ref{bulk}(c) and \ref{bulk}(d). Namely, the intersection at $\Delta>0$ is an EP3 while the one at $\Delta<0$ is a PEP2, with braid invariants $[b_{\Gamma_1}]=[\sigma_3\sigma_4]$ and $[b_{\Gamma_2}]=[\sigma_3\sigma_5]$ respectively, where the reference loops $\Gamma_{1,2}$ are shown in figure \ref{bulk}(a). To understand how the difference comes about, consider the braid invariant on a loop parametrized by the intermediate value of $\Delta=\Delta_m$, which tunes between the reference loops $\Gamma_{1,2}$. The point is that there are topologically distinct such loops, distinguished by whether they enclose the extra (blue) exceptional line or not (compare the loops $\Gamma_1'$ and $\Gamma_2'$ in figure \ref{bulk}(b)). According to the NACR, the braid invariant computed from the loop $\Gamma_1'$ is equal to the one computed from the loop $\Gamma_1$, and similarly for $\Gamma_2$ and $\Gamma_2'$. However, if the braid invariant of $\Gamma_2'$ is $b_{\Gamma_2'} = b_X b_Y$, then the invariant computed from $\Gamma_1'$ is (fig. \ref{bulk}(b))
\begin{equation}
    b_{\Gamma_1'} = b_Z^{-1} b_X b_Z b_Y. \label{b_Gamma1'}
\end{equation}
If $b_Z$ is non-commuting with $b_X$ and $b_Y$, namely $b_{X(Y)}b_Z \neq b_Z b_{X(Y)}$, then $b_{\Gamma_1'}$ and $b_{\Gamma_2'}$ are distinct braid invariants. In our parametrization, 
\begin{equation}
    b_X = \sigma_3,\ b_Y=\sigma_5,\ b_Z = \sigma_4, \label{braid}
\end{equation}
at some specific value of $\Delta$. Subsituting (\ref{braid}) into (\ref{b_Gamma1'}), we arrive at the (topologically equivalent) braid invariant of the EP3. Thus in this example, the initially commuting ELs ends up non-commuting, resulting in HOEPs, due to their non-Abelian braiding with other ELs in parameter space. As this case illustrates, the braiding structure of ELs follows naturally from keeping track of the eigenvalue braiding invariant associated to a given loop around the EL. As the loop is moved along the EL, it might need to intersect other ELs if these are linked, leading to a modified braid invariant \cite{ding2016emergence,doppler2016dynamically}.

\begin{figure}
\centering
\includegraphics[width=1\linewidth]{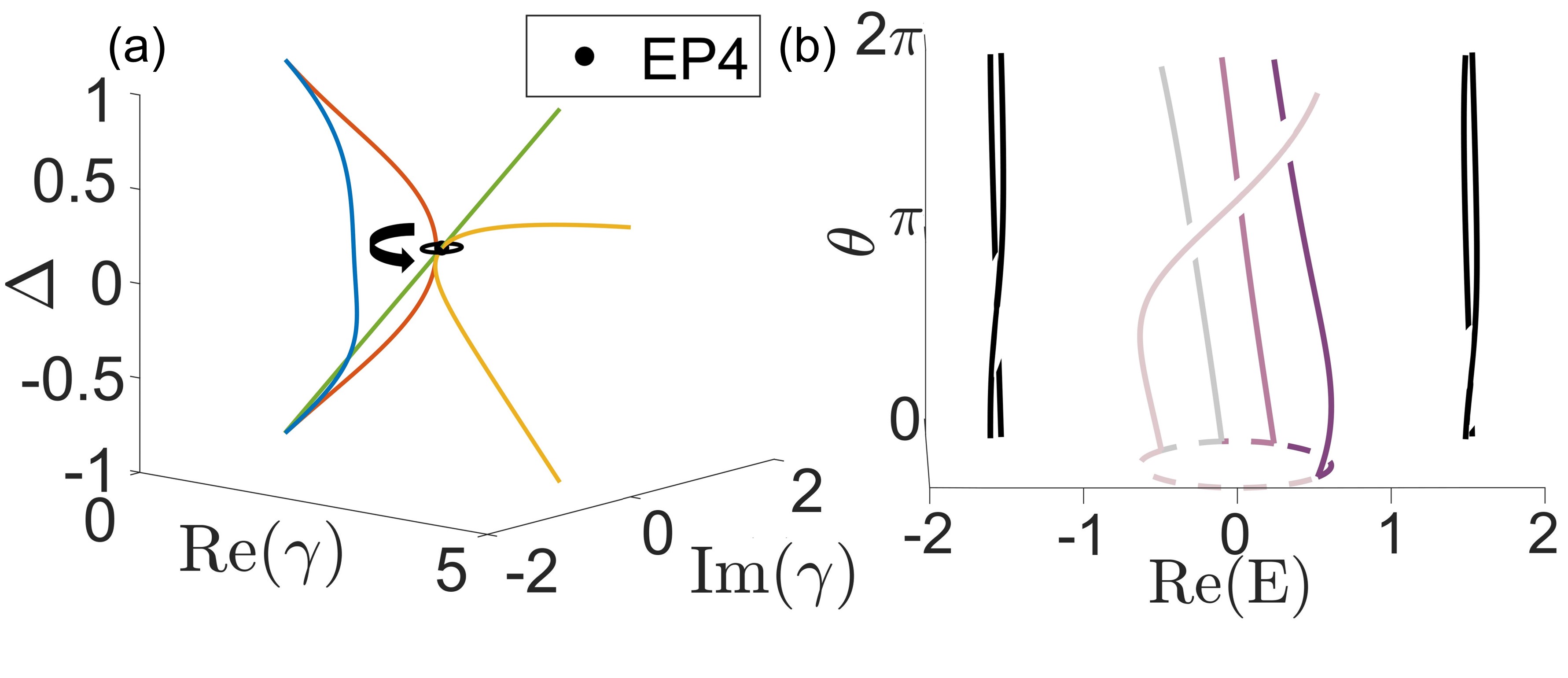}
\caption{\justifying Eigenvalue braiding and ELs for the SSH chain with a midpoint defect ($s=\frac{N}{2}+1$), where $N=4,s=3$. (a) Closed paths encircling the EP4 (black dot). (b) Eigenvalue braiding near the EP4. The path is $\gamma = 2.28 + 0.2i e^{i\theta}$ with $\theta:0 \rightarrow 2\pi$ and $\Delta = 0.14$.} 
\label{mid}
\end{figure}

A special case appears when the sublattice length $N=2m$ is even and the defect location $2s-1$  corresponds to the mid-point, $s=m+1$. At the intersection of the green and yellow curves in figure \ref{mid}(a) we find an EP4, with braiding $[b_{\rm EP4}] =  [\sigma_1 (\sigma_3 \sigma_4 \sigma_5) \sigma_7] $ (see figure \ref{mid}(b)). An interesting aspect of this result is that typically one only expects an EP4 to appear at the intersection of at least three non-commuting EP2s. Here, the EP4 appears because the green line is a very special case -- the stable PEP2, as we mentioned in equation (\ref{eq:charac_evenN}), where a small perturbation of $\Delta$ does not split it into EP2s. Along the green exceptional line, the whole spectrum coalesces in pairs. Thus the invariant on the green line is $[\sigma_1 \sigma_3 \sigma_5 \sigma_7]$, and its merging with the non-commuting yellow line $[\sigma_4]$ leads to the EP4 braiding $b_{\rm EP4} =  [\sigma_1 (\sigma_3 \sigma_4 \sigma_5) \sigma_7]$, with the additional braiding of two separate pairs of levels, $[\sigma_1]$ and $[\sigma_7]$.

The PEP2s do not always correspond to the appearance of an EP4. For example, when $N$ is odd, the ELs diagram in \ref{ELs}(d) has a similar structure as the diagram for the ELs in  the bulk-site defect case, and their intersections also lead to an EP3. This can also be explained by equation (\ref{b_Gamma1'}). In this case,
\begin{equation}
    b_X = \sigma_4,\ b_Y=\sigma_6,\ b_Z = (\sigma_1\sigma_3)\sigma_5(\sigma_7\sigma_9).
\end{equation}
Here, only $\sigma_5$ is the non-commuting part with $b_X$ and $b_Y$. Thus such a PEP2 leads to an EP3 rather than EP4. We will go back to the reason behind this in section \ref{roleofedge}.

In the dimerized limits $\Delta = \pm 1$, the ELs converge to several fixed points. One fixed point is the Hermitian degeneracy at $\gamma=0$, where the EPs of opposite chirality, or equivalently with a braid invariant and its inverse, annihilate \cite{krol2022annihilation}. In the bulk defect case, the ELs also converge at the EP of the dimerized chain $\gamma = 4$. In the boundary defect case, a special fixed point appears at $\Delta = 1$, $\gamma = \pm 2i$, where the boundary site becomes isolated. Interestingly, not all ELs are symmetric under sign change of $\Delta$. In fact, the HOEPs only appear in the topological phase $\Delta>0$. As we now discuss, the formation of HOEPs can be traced back to the topological edge state.

\begin{figure}

\centering
\includegraphics[width=1\linewidth]{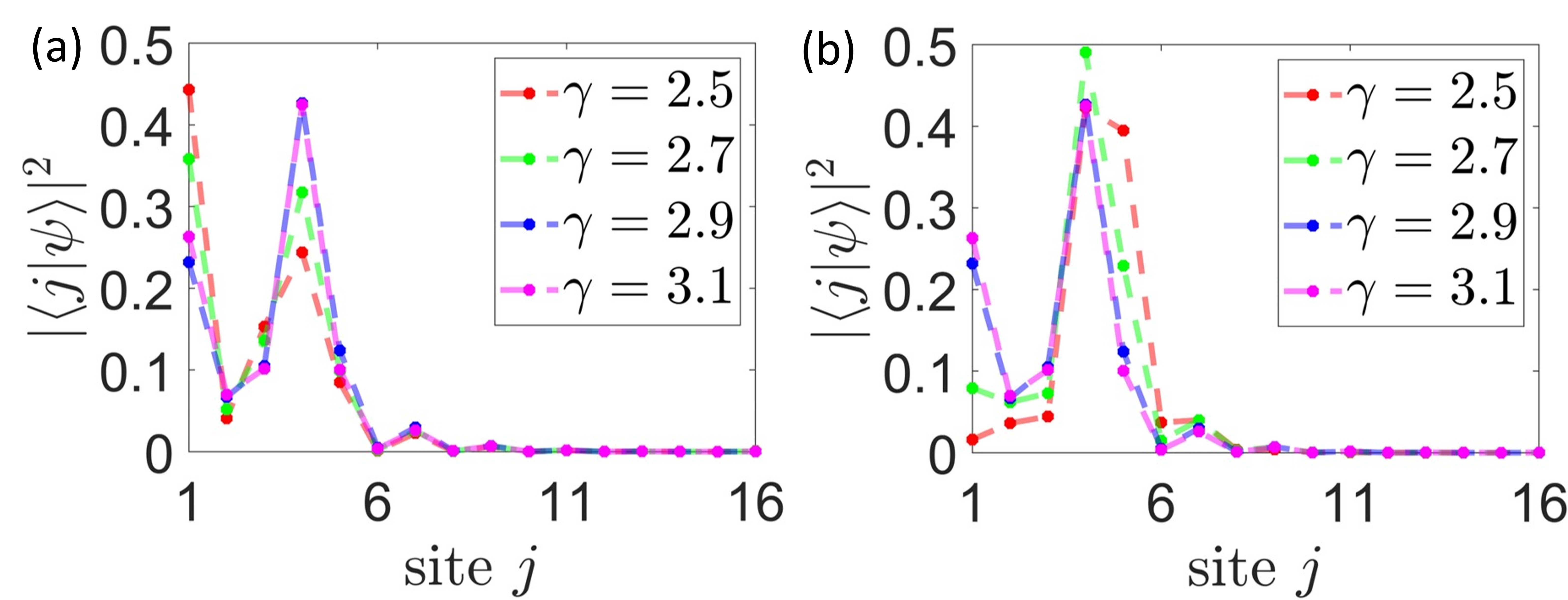}
\caption{\justifying {Delocalization of the edge states and defect states near $\gamma_{\rm mixed}\approx 2.82$. (a) Wavefunction profile of left edge states. (b) Wavefunction profile of defect state. In both figures, $2N=16,s=3,\Delta=0.3$. }}
\label{delocal}
\end{figure}

\section{The role of edge states}\label{roleofedge}

\begin{figure*}
\centering
\includegraphics[width=1\linewidth]{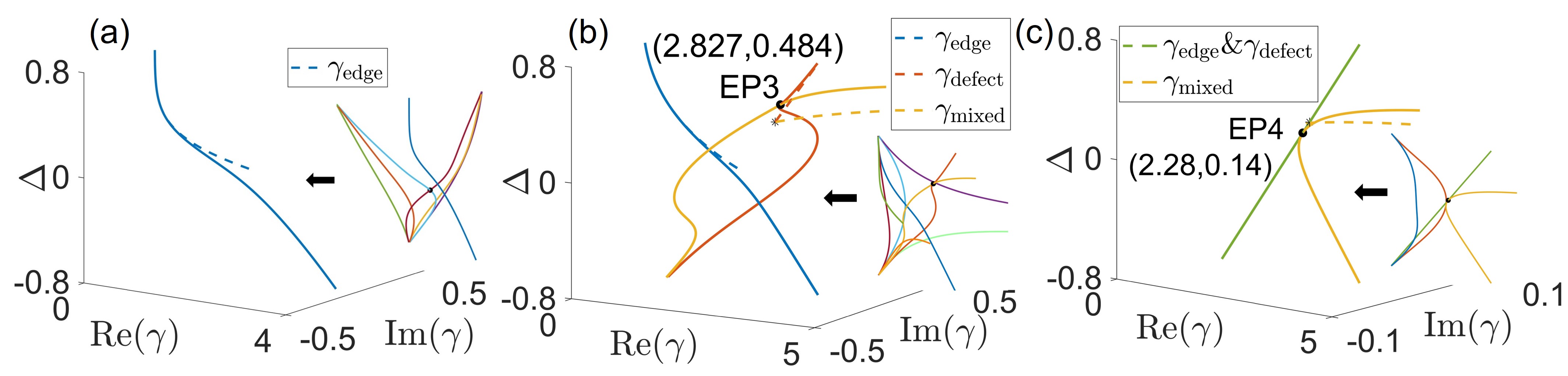}
\caption{\justifying Comparison between the predicted and the analytical values of ELs when the defect is located at (a) the boundary $s=1$, (b) a generic bulk site $s=2$ and (c) midpoint $s=3$. For all figures, $N=4$ and the inset is the complete structure of ELs in each case. }
\label{EL_app}
\end{figure*}

In the topological phase $\Delta>0$, the spectrum is gapped and to leading order one can study the effect of the non-Hermitian defect on the in-gap states separately.
The projection of $H$ onto the edge states gives
\begin{equation}
    H_{\rm eff} = \left(
    \begin{array}{cc}
       \langle L|H|L\rangle  & \langle L|H|R\rangle \\
       \langle R|H|L\rangle  &  \langle R|H|R\rangle      
    \end{array}
    \right)
    \equiv
    \left(
    \begin{array}{cc}
       \mathcal E_L  & J \\
       J^*  &  \mathcal E_R      
    \end{array}
    \right).
\end{equation}
The overlap $J\propto r^{-N}$ is responsible for the exponentially small energy splitting in the Hermitian case, $\mathcal{E}_L\propto i\gamma r^{-2s}$ is proportional to the squared amplitude of the edge state at the defect site $2s-1$, and $\mathcal{E}_R=0$ for our choice of placing the defect at an odd site. Here, $r^{-1}=\frac{1-\Delta}{1+\Delta}=e^{-\xi}$ parametrizes the localization length $\xi$ of the edge states. This two-level Hamiltonian has an exceptional point at $\mathcal{E}_L^2=4|J|^2$. In the dissipative region $\text{Re}(\gamma)>0$ this gives the exceptional point
\begin{equation}
    \gamma_{\rm edge} = 2(1+\Delta)r^{2s-2-N}.\label{eq:gammaedge}
\end{equation}
The defect also leads to localized states in the bulk. Their wavefunctions are given by
\begin{equation}
    \psi_n = 
\left\{
\begin{aligned}
& A\exp(-\kappa|l-s|), \hspace{1cm} n = 2l \\
& B\exp(-\kappa|l-s+1|), \hspace{.4cm}n = 2l+1
\end{aligned}
\right.,
\label{wave}
\end{equation}
where the localization length $\kappa^{-1}$ and the corresponding eigenenergies $\lambda$ are given by
\begin{align}
    & \lambda^2 = 2(1-\Delta^2) \cosh(2\kappa) + 2(1+\Delta^2), \\
    &  \cosh(2\kappa) = \frac {-\gamma^2 \pm \sqrt{(\gamma^2-4(1-\Delta)^2)(\gamma^2-4(1+\Delta)^2)}}{4 (1-\Delta^2)},\nonumber
\end{align}
which have degeneracies $\gamma_{\pm} = 2(1 \pm \Delta)$ in the dissipative regime. The condition for localized states $\text{Re}(\kappa)>0$ singles out the exceptional point
\begin{equation}
    \gamma_{\rm defect} = 2(1+\Delta),\label{eq:gammaimpu}
\end{equation}
where the two defect states coalesce. A third possibility is the coalescence of one of the defect states with one of the edge states. When $\gamma \geq \max({\gamma_{\rm edge}, \gamma_{\rm defect}}),$ the four eigenenergies become fully imaginary, and one can estimate that such a mixed coalescence happens at
\begin{equation}
    \gamma_{\rm mixed} =(1+\Delta)r^{s-1}\sqrt{1-r^{-2N}}.
    \label{eq:gammamixed}
\end{equation}
The coalescing wavefunctions in this case are shown in figures \ref{delocal}(a), \ref{delocal}(b).

Equations (\ref{eq:gammaedge}), (\ref{eq:gammaimpu}), (\ref{eq:gammamixed}) approximate the ELs involving localized in-gap states, and are plotted as dashed lines in figures {\color{red} \ref{EL_app}(a)-\ref{EL_app}(c)}. HOEPs can appear when they intersect at the saturation of the $\gamma_{\rm mixed}\geq \text{max}(\gamma_{\rm edge},\gamma_{\rm defect})$ bound. For $s<\frac{N}{2}+1$, $\gamma_{\rm defect}(\Delta)>\gamma_{\rm edge}(\Delta)$, and an EP3 appears at $\gamma_{\rm mixed}(\Delta_{\rm EP3}) = \gamma_{\rm defect}(\Delta_{\rm EP3})$, whose solution $(\gamma_{\rm EP3},\Delta_{\rm EP3})$ is shown in figure {\color{red} \ref{EL_app}(b)} and is very close to the numerical value of the observed EP3. It corresponds to the simultaneous coalescence of the two defect states and the left edge state. For $s>\frac{N}{2}+1$, $\gamma_{\rm edge}(\Delta)>\gamma_{\rm defect}(\Delta)$, and an EP3 appears due to the simultaneous coalescence of the two edge states and one defect state at $\gamma_{\rm mixed}(\Delta_{\rm EP3}) = \gamma_{\rm edge}(\Delta_{\rm EP3})$. Note that this asymmetry is due to our choice of placing the defect at an odd site $2s-1$. Instead, if the defect is located at $2s$, then the situation is reversed and, for $s>\frac{N}{2}+1$, an EP3 is formed due to the simultaneous coalescence of the two defect states and the right edge state.

When $N$ is even, there is a special case at the midpoint $s = \frac{N}{2} + 1$, for which $\gamma_{\rm edge}(\Delta)=\gamma_{\rm defect}(\Delta)$, the stable PEP2. The four localized states coalesce, generating an EP4, at $\gamma_{\rm mixed}(\Delta_{\rm EP4}) = \gamma_{\rm defect}(\Delta_{\rm EP4}) = \gamma_{\rm edge}(\Delta_{\rm EP4})$. The solution is once again close to the numerically computed value, as shown in figure {\color{red} \ref{EL_app}(c)}.

Finally, when $s=1$, no additional localized states are generated besides the edge states. Thus we only have $\gamma_{\rm edge}(\Delta)$, shown in figure {\color{red} \ref{EL_app}(a)}, and no HOEP, but instead PEP2s formed from the coalescence of two pairs of scattering states. For more details of derivations of Equations (\ref{eq:gammaedge}), (\ref{eq:gammaimpu}), (\ref{eq:gammamixed}) and the validity of the approximation method, see Appendix \ref{EPofstates}.

Notice that the results for cases of boundary defect and of mid-point defect are somewhat fined-tuned. In the thermodynamic limit $N\to\infty$, the exceptional degeneracy corresponding to the edge states $\gamma_{\rm edge}\to 0$ disappears into their exact Hermitian degeneracy, and the EP2 in the boundary defect case and the EP4 in centered defect case also disappear.
However, $\gamma_{\rm mixed}$ remains finite in the limit of a half-infinite chain where the distance of the defect to the boundary, $2s-1$, is kept fixed. Thus in this case the EP3 survives the thermodynamic limit and can be detected at finite loss rate $\gamma_{\rm EP3}$. It is worth emphasizing that this phenomenology is distinct from the anomalous bulk-boundary correspondence of fully non-Hermitian SSH models \cite{Yao2018,Lee2016,Xiong2018,Kunst2018} at which the boundary modes form an EP2 in the thermodynamic limit \cite{Kunst2018} due to the non-Hermitian skin effect \cite{Okuma2023,Lin2023}. 

\section{Discussion}\label{sec:discussion}

Recall that, from our counting of resultants, a degeneracy of multiplicity $3$ typically requires tuning $2(3-1)=4$ real parameters, and a degeneracy of multiplicity $4$ requires tuning $2(4-1)=6$ real parameters. However, we found that in the simple model of the SSH chain perturbed by a localized lossy defect, both an EP3 and an EP4 appear in the 3D space $(\Delta,{\rm Re}(\gamma),{\rm Im}(\gamma))$. The reason why this happens is two-fold. First, the SSH chain has topological edge states. While it was previously realized that non-Hermiticity can have the independent effects of (a) delocalizing the edge states \cite{cheng2022competition, zhu2021delocalization} and (b) leading to EPs due to the coalescence of non-Hermitian defect states \cite{burke2020non}, we find that in fact (a) and (b) can collaborate to create HOEPs in our model by the simultaneous coalescence of two defect states and one edge state. This point is enough to explain the EP3. Second, when the chain length $2N$ is a multiple of $4$, ie. $N$ is even, and the defect is located at the midpoint $s=\frac{N}{2}+1$, stable PEP2s appear: a special EL where the full spectrum coalesces in pairs. Together with the first point, it leads to the simultaneous coalescence of the two defect states and the two edge states, forming the EP4. These two points reveal topological features of the model which only become apparent in the multiband non-Hermitian setting. In parameter space, they translate into the special intersections of non-commuting ELs, which manifest their non-Abelian braiding properties.
Our example opens up a new window of exploration: what types of non-Abelian, non-Hermitian topology can arise from the interplay between local defects and boundary states in different symmetry classes? This question naturally branches out in three sets of intriguing problems: first, for the case of (symmetry-protected) Hermitian topological phases \cite{HasanKane,QiZhang,RevModPhys.88.035005,Asb_th_2016} with non-Hermitian defects as studied here in a minimal model. Second, non-Hermitian symmetry-protected phases which already in the bulk exhibit non-Abelian topological features \cite{Yang2023} which may be further enriched by the interplay between boundaries and defects. Third, different dimensionality and distributions of defects, which may further induce unconventional phenomena and higher exceptional points.

{\it Acknowledgments --}  This work was sponsored by Pujiang Talent Program 21PJ1405400, Jiaoda 2030 program WH510363001-1, and the Innovation Program for Quantum Science and Technology Grant No.2021ZD0301900. EJB is supported by the Swedish Research Council (VR, grant 2018-00313), the Wallenberg Academy Fellows program (2018.0460) of the Knut and Alice Wallenberg Foundation, and the G\"oran Gustafsson Foundation for Research in Natural Sciences and Medicine.

\appendix

\section{Characteristic polynomial of the SSH chain with a lossy defect}\label{charac}

The characteristic matrix $\lambda I-H$ of the SSH Hamiltonian with lossy defect has a tridiagonal form,
\begin{equation}
    \lambda I - H = 
\left(
\begin{matrix}
a_1 & b_1 &   & & \\
c_1 & a_2 & b_2 & & \\
 & c_2 & a_3 & b_3 & \\
 &  &  &\ddots & b_{n-1} \\
 & & & c_{n-1} & a_n
\end{matrix}
\right),\label{eq:tridiagonal}
\end{equation}
with coefficients
\begin{equation}
\begin{aligned}
    b_n = \, & c_n = \left \{
    \begin{aligned}
        & 1-\Delta  &  n = 2l-1,\, \forall l\\
        & 1+\Delta &  n = 2l,\, \forall l
    \end{aligned}
    \right., \\
    & a_n = \left \{
    \begin{aligned}
        & \lambda & n \neq 2s-1 \\
        & \lambda + i \gamma \quad  & n = 2s-1
    \end{aligned}
    \right. .
\end{aligned}
\end{equation}
The determinant $P_n$ of a tridiagonal matrix (\ref{eq:tridiagonal}) satisfies the recurrence relation
\begin{equation}
    P_n = a_n P_{n-1} - b_{n-1} c_{n-1} P_{n-2},
\end{equation}
which can be proved using Laplace's expansion theorem. In the absence of defect, this recurrence relation becomes
\begin{equation}
    \left \{
    \begin{aligned}
        & P_{2l} = \lambda P_{2l-1} - (1-\Delta)^2 P_{2l-2}\\
        & P_{2l+1} = \lambda P_{2l} - (1+\Delta)^2 P_{2l-1}
    \end{aligned}
    \right. ,\label{eq:recursion}
\end{equation}
and it can be solved by the ansatz $P_{2l} = A z^{l}, P_{2l+1} = C z^{l}$. Substituting the ansatz in (\ref{eq:recursion}) one finds two roots,
\begin{equation}
     z_\pm = (1-\Delta^2) (Q \pm \sqrt{Q^2 - 1}),
\end{equation}
where $Q = \frac{\lambda^2 - 2(1+\Delta^2)}{2(1-\Delta^2)}$. For simplicity, we define $z = Q + \sqrt{Q^2 - 1}=(Q - \sqrt{Q^2 - 1})^{-1}$, so that $P_{2l} = (1-\Delta^2)^l (A z^{l} + B z^{-l})$ and $P_{2l+1} = (1-\Delta^2)^l (C z^{l} + D z^{-l})$ where, by (\ref{eq:recursion}), only two of the $A,B,C,D$ coefficients are independent and should be determined by initial conditions. It is convenient, however, to treat them as independently determined by $P_0$, $P_1$, $P_2$ and $P_3$. We find
\begin{equation}
\begin{aligned}
    & P_{2l} = (1-\Delta^2)^{l-1}(P_2 U_{l-1} - (1-\Delta^2) P_0 U_{l-2}), \\
    & P_{2l+1} = (1-\Delta^2)^{l-1}(P_3 U_{l-1} - (1-\Delta^2) P_1 U_{l-2}).
\end{aligned}\label{eq:recursioninitial}
\end{equation}
Here, $U_n= U_n(Q)$ is the $n$-th Chebishev polynomial of the second kind, and in deriving this formula we used that $Q=\frac{z+z^{-1}}{2}$ and the property
\begin{equation}
    U_n\left(\frac{z+z^{-1}}{2}\right)=\frac{z^{n+1}-z^{-n-1}}{z-z^{-1}}. \label{Cheby_prop}
\end{equation}
Additional useful properties of Chebyshev polynomials are \cite{mason2002chebyshev}:
\begin{equation}
    \begin{aligned}
        & U_n(x) = 2x U_{n-1} (x) - U_{n-2} (x) ,\\ 
        & U_n = U_m U_{n-m} - U_{m-1}U_{n-m-1} ,\\
        & \frac {dU_n(x)}{dx} = \frac{(n+1)U_{n+1} - nxU_{n}}{1-x^2} .
    \end{aligned}
    \label{Cheby}
\end{equation}
For the unperturbed SSH chain, $P_0^{(0)} = 1$, $P_1^{(0)} = \lambda$, $P_2^{(0)} = \lambda^2 - (1-\Delta)^2$, $P_3^{(0)} = \lambda (\lambda^2 - 2(1+\Delta^2))$, so that (\ref{eq:recursioninitial}) gives
\begin{equation}
\begin{aligned}
    & P_{2l}^{(0)} = (1-\Delta^2)^{l}\left(U_{l} + rU_{l-1}\right), \\
    & P_{2l+1}^{(0)} = (1-\Delta^2)^{l}\lambda U_l ,
\end{aligned}\label{eq:SSHrecursion}
\end{equation}
where $r=\frac{1+\Delta}{1-\Delta}$. Thus the characteristic polynomial of the SSH Hamiltonian with $2N$ sites is
\begin{equation}
    P_{2N}^{(0)} = (1-\Delta^2)^{N}\left(U_{N} + rU_{N-1}\right).
\end{equation}
The presence of a lossy defect at site $2s-1$ changes the recurrence at a single site, so that
\begin{equation}
\begin{aligned}
    P_{2s-1} &= (\lambda + i\gamma)P_{2s-2} - (1+\Delta)^2 P_{2s-3}\\
    &= P_{2s-1}^{(0)}+i\gamma P_{2s-2}^{(0)},
\end{aligned}
\end{equation}
and likewise
\begin{equation}
    P_{2s} = P_{2s}^{(0)}+i\gamma\lambda P_{2s-2}^{(0)}.
\end{equation}
Now we can take $P_{2s-2}$ and $P_{2s}$ as initial conditions in (\ref{eq:recursioninitial}),
\begin{equation}
\begin{aligned}
    P_{2N} &= (1-\Delta^2)^{N-s}(P_{2s}U_{N-s}-(1-\Delta^2)P_{2s-2}U_{N-s-1}) \\
    &= P_{2N}^{(0)}+i\gamma\lambda(1-\Delta^2)^{N-s}P_{2s-2}^{(0)}U_{N-s},
\end{aligned}
\end{equation}
and finally
\begin{equation}
\begin{aligned}
    P_{2N} = & (1-\Delta^2)^{N} \times \\ 
    & \left[(U_N + r U_{N-1}) + \frac{i\gamma \lambda}{1-\Delta^2} (U_{s-1} + r U_{s-2})U_{N-s}\right].
\end{aligned}
\label{CP}
\end{equation}
Note that this expression also holds in the case of defect at the boundary, $s=1$, if we extend the definition of Chebyshev polynomials by $U_{-1} = 0$.

\section{Conditions for line degeneracies} \label{ld}

To compute the resultants, we first disregard the constant prefactor and redefine
\begin{equation}
    p(\lambda) = U_{N} + rU_{N-1}+\frac {i\gamma\lambda}{1-\Delta^2}\left(U_{s-1} + rU_{s-2}\right)U_{N-s}. \label{p}
\end{equation}
We expand it using
\begin{equation}
    U_n = z^n + z^{n-2} + \cdots + z^{-n+2} + z^{-n},
\end{equation}
and
\begin{equation}
    z^n + z^{-n} = \left\{
    \begin{aligned}
        & 2 \sum_{l=0}^{m} C_n^{2l} (Q^2-1)^l Q^{2(m-l)}, \quad n=2m \\
        & 2 \sum_{l=0}^{m} C_n^{2l} (Q^2-1)^l Q^{2(m-l)+1}, \quad n=2m+1, \\
    \end{aligned}
    \right.
\end{equation}
where $C_n^m=\frac{n!}{m!(m-n)!}$ is the binomial number, so that the characteristic polynomial is the standard form. It is useful to redefine the variable $\lambda^* = \frac{\lambda}{\sqrt{1-\Delta^2}}$, so that $Q = \frac{1}{2}((\lambda^*)^2 - r - r^{-1})$, and introduce the notation $a = \frac{\gamma^2}{1-\Delta^2}$. Then the coefficients of the characteristic polynomial 
\begin{equation}
    p(\lambda^*) = U_{N} + rU_{N-1}+i \sqrt{a} \lambda^*\left(U_{s-1} + rU_{s-2}\right)U_{N-s}
\end{equation}
are all expressed in terms of $r$ and $a$. Subtituting the standard form of the polynomials into equation (\ref{res}) and using the Symbolic Math Toolbox of MATLAB R2023b, we further obtain the condition for line degeneracies with different length of the chain and the defect locations.

\section{Paired degeneracy} \label{paireddeg}

We give a detailed derivation for the paired degeneracies in (\ref{eq:charac_evenN}) and (\ref{eq:charac_oddN}). We begin with (\ref{eq:charac_evenN}) and set $N=2m$ and $s=m+1$. Substituting into (\ref{p}), 
\begin{equation}
    p = U_{2m} + rU_{2m-1}+\frac {i\gamma\lambda}{1-\Delta^2}\left(U_{m} + rU_{m-1}\right)U_{m-1}.
\end{equation}
Using the properties of Chebyshev polynomials (\ref{Cheby}), one can show that
\begin{equation}
    \begin{aligned}
        U_{2m} &= U_m^2-U_{m-1}^2,\\
        U_{2m-1}&=2 U_m U_{m-1}-2QU_{m-1}^2,
    \end{aligned}
\end{equation}
from which it follows that
\begin{equation}
\begin{aligned}
    p = & \left(2r+\frac{i\gamma\lambda}{1-\Delta^2}\right)U_mU_{m-1} + \\
    & U_m^2 + \left(r^2 - \frac{\lambda^2}{(1-\Delta)^2} + \frac{i\gamma\lambda r}{1-\Delta^2}\right)U_{m-1}^2.
\end{aligned}
\end{equation}
When $\gamma=2(1+\Delta)$, this polynomial is indeed a perfect square
\begin{equation}
    p = \left(U_m + \frac{i\lambda+ 1+\Delta}{1-\Delta}U_{m-1}\right)^2.
\end{equation}

For (\ref{eq:charac_oddN}), we set $N=2m-1$ and $s=m$, so that
\begin{equation}
    p = U_{2m-1} + rU_{2m-2}+\frac {i\gamma\lambda}{1-\Delta^2}\left(U_{m-1} + rU_{m-2}\right)U_{m-1}.
\end{equation}
Similarly, expanding the Chebyshev polynomials,
\begin{equation}
\begin{aligned}
    p = &\left(\frac{i\gamma\lambda r}{1-\Delta^2} - 2\right)U_{m-1}U_{m-2} + \\
    & \left(2Q+\frac{i\gamma\lambda}{1-\Delta^2} + r \right)U_{m-1}^2 - r U_{m-2}^2.
\end{aligned}
\end{equation}
When $\gamma = 2(1-\Delta)$, the polynomial is again a perfect square
\begin{equation}
    p = r\left(\frac{\lambda+i(1-\Delta)}{1+\Delta}U_{m-1} + i U_{m-2}\right)^2.
\end{equation}

\section{Exceptional points from localized eigenstates} \label{EPofstates}

\subsection{ Exceptional points from edge states}

The Hermitian SSH chain in thermodynamic limits possesses two degenerate edge states, namely the left and the right edge states \cite{Asb_th_2016}
 \begin{equation}
     |L\rangle = \sum_{l=1}^N a_{l} |2l-1\rangle, \quad 
     |R\rangle = \sum_{l=1}^N b_{l} |2l\rangle
     \label{edge_states}
 \end{equation}
where $a_l = (-1)^{l-1} a_1 r^{-(l-1)} ,\quad b_l = (-1)^{N-l} b_N r^{-(N-l)}$, with normalized amplitudes given by
\begin{equation}
    |a_1|^2 = |b_N|^2 = \frac{1-r^{-2}}{1-r^{-2N}}.
\end{equation}
The leading effects of defect on the edge states are captured by the effective two-level Hamiltonian
\begin{equation}
    H_{\rm eff} = \left(
    \begin{array}{cc}
       \langle L|H|L\rangle  & \langle L|H|R\rangle \\
       \langle R|H|L\rangle  &  \langle R|H|R\rangle  
    \end{array}
    \right)
    \equiv
    \left(
    \begin{array}{cc}
       \mathcal E_L  & J \\
       J^*  &  \mathcal E_R      
    \end{array}
    \right),
    \label{effectiveH}
\end{equation}
where
\begin{equation}
    \mathcal E_L = -i\gamma|a_s|^2, \mathcal E_R = 0, J = (-1)^{N} (1-\Delta) a_1^*b_N r^{-(N-1)}.
\end{equation}
Therefore, the eigenvalues of the effective Hamiltonian are 
\begin{equation}
    \lambda_{\rm edge,\pm} = \frac 12 ( \mathcal E_L \pm \sqrt{4|J|^2 +  \mathcal E_L^2})
\end{equation}
and the corresponding eigenstates are
\begin{equation}
    \psi_\pm =\frac {1}{A}\left(J, \frac 12 (\mathcal E_L \pm \sqrt{ 4|J|^2 + \mathcal E_L^2})
    \right)^T,
\end{equation}
where $A$ is a normalization factor,
\begin{equation}
    A = \left\{
    \begin{aligned}
        & \sqrt{2} |J| & \gamma \leq \gamma_{\rm edge} \\
        & \frac 12\sqrt{2 i \mathcal E_L (i \mathcal E_L \pm \sqrt{ -4|J|^2 - \mathcal E_L^2}}) &\gamma > \gamma_{\rm edge}
    \end{aligned}
    \right. .
\end{equation}
The EPs lie at the coalescence of the eigenenergies, $4|J|^2 + \mathcal E_L^2 = 0$, which correspond to
\begin{equation}
    \gamma_{\rm edge}(\Delta) = \pm 2(1+\Delta)r^{2s-2-N}. \label{eq:edgeEP}
\end{equation}
Since we restrict the defect to be dissipative, only the positive solution is kept.

At finite length, the edge states overlap and give rise to the symmetric/anti-symmetric states $\frac{1}{\sqrt{2}}(|L\rangle\pm|R\rangle)$. When $\gamma \leq \gamma_{\rm edge}$, two eigenstates are
\begin{equation}
    \psi_\pm = \frac {1}{\sqrt{2}}\left(1,e^{-i \varphi_\pm(\gamma)}
    \right)^T,
\end{equation}
where $\varphi_\pm(\gamma) = \arctan \frac{\gamma}{\gamma_{\rm edge}^2  - \gamma^2}, \pi - \arctan \frac{\gamma}{\gamma_{\rm edge}^2  - \gamma^2}$. At the EP the Hamiltonian is defective with only one eigenstate $\frac{1}{\sqrt{2}}(|L\rangle+e^{-i\frac \pi 2} |R\rangle)$.

\begin{figure}
\centering
\includegraphics[width=1\linewidth]{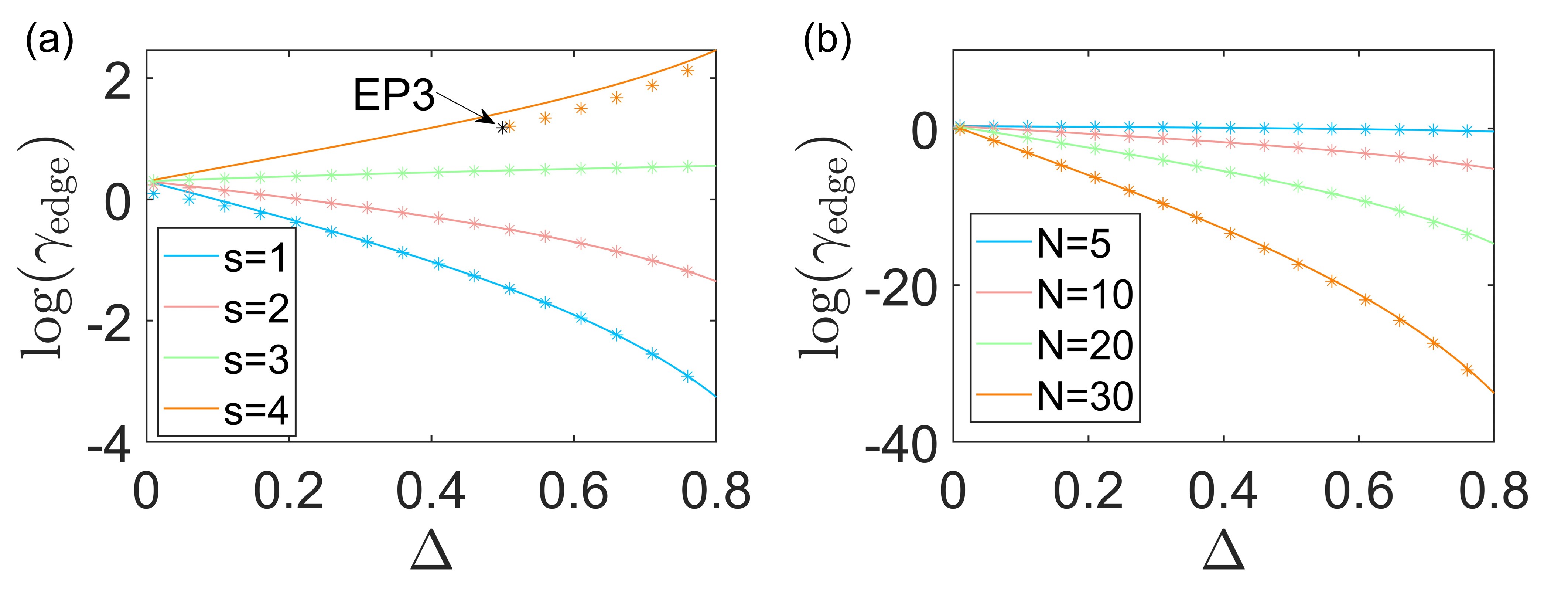}
\caption{\justifying { Comparison between the $\gamma_{\rm edge}$ from (\ref{eq:edgeEP}) (lines) and the numerical results (dots) with (a) different defect locations $s$ and a fixed length $2N = 8$. (b) different chain lengths and a fixed defect location $s=3$.}}
\label{edgeEP}
\end{figure} 

For completeness, we estimate the mixing of edge states with scattering states, which was neglected above. Since the defect is located on the odd lattice sites, then to first order only the left edge state is modified,
\begin{equation}
    \begin{aligned}
        & |L'\rangle = |L\rangle + \frac {i\gamma a_s}{N} \sum_{l=1}^N \sum_{k} e^{ik(s-l)-i\phi(k)} |2l\rangle + O(\gamma^2), \\
        & \phi(k) = \tan^{-1} \frac {(1-\Delta) \sin k}{ 1+\Delta+ (1-\Delta) \cos k},
    \end{aligned}
\end{equation}    
where 
\begin{equation}
    k = \frac{2\pi n}{N-2},\quad (n=0,1,\cdots,N-3)
\end{equation}
is the lattice momentum with discretized values at finite length.
Thus, the two-level approximation is valid for $\gamma|a_s| \ll 1$. As a lower bound, the emergence of the edge EPs requires $\gamma_{edge}|a_s| \propto r^{s-N} \ll 1$. This is reflected in figures \ref{edgeEP} and \ref{defectEPb}. Interestingly, when the defect is located at the right side of the midpoint, namely $s>N/2+1$, the numerical results are cut off at a range. This is because of the extension of the EP values out of the plane into complex region, after two EPs coalesce at a HOEP, in this case, an EP3. 

\subsection{Exceptional points from defect states}

The defect states have the form
\begin{equation}
    \psi_l = 
\left\{
\begin{aligned}
& A\exp(-\kappa|l-s|) \\
& B\exp(-\kappa|l-s+1|)
\end{aligned}
\right. \quad l=1,\cdots,N,
\label{impu}
\end{equation}
where $\kappa$ is the localization rate with $\Re(\kappa)>0$. Assuming an infinite chain, such an ansatz gives an eigenstate,
\begin{equation}
    \sum_j H_{ij} \psi_j = \lambda \psi_j \label{Ham}
\end{equation}
where 
\begin{equation}
\begin{aligned}
    H_{ij} \equiv \langle i |H|j\rangle = &  -(1+(-1)^i)\Delta)\delta_{i,j+1} -  \\ 
    & (1+(-1)^j)\Delta)\delta_{i+1,j} - i\gamma\delta_{i,m}\delta_{j,m}
\end{aligned}
\label{Sch}
\end{equation}
is the Hamiltonian matrix. Since the chain is assumed infinite, different choices of the sublattice sites are equivalent, by translating the defect along the chain. For simplicity, we set $2s-1=0$,
\begin{equation}
\left\{
    \begin{aligned}
    & \lambda \psi_{2l-1} = -(1-\Delta)\psi_{2l-2} -(1+\Delta) \psi_{2l} \quad (l\in \mathbb Z), \\
    & \lambda \psi_0 = -(1+\Delta)\psi_{-1} -(1-\Delta)\psi_1 - i\gamma\psi_0, \\
    & \lambda \psi_{2l} = -(1+\Delta)\psi_{2l-1} -(1-\Delta)\psi_{2l+1} \quad (l\in \mathbb Z,l\neq 0). 
\end{aligned}
\right. \label{equ_sys}
\end{equation}
Only three independent equations in (\ref{equ_sys}) are left after substituting (\ref{impu}), which give
\begin{equation}
\begin{aligned}
    & \lambda^2 = 2(1-\Delta^2)e^{-2\kappa} + 2(1+\Delta^2) - i\gamma \lambda, \\
    & \lambda^2 = (1-\Delta^2)(e^{2\kappa}+e^{-2\kappa}) + 2(1+\Delta^2),
\end{aligned}
\label{lambda}
\end{equation}
and the relation for the coefficients $A$ and $B$ is
\begin{equation}
    Be^{-\kappa} = -(1-\Delta)e^{-2\kappa} - (1+\Delta)A.
\end{equation}
Let $x = e^{2\kappa}, y = x+1/x$. We can determine $y$ from (\ref{lambda}),
\begin{equation}
    y^2 + \frac {\gamma^2}{(1-\Delta^2)} y + \frac {2\gamma^2 (1+\Delta^2) }{(1-\Delta^2)^2}-4 = 0,
\end{equation}
with the solutions
\begin{equation}
    y_\pm = \frac {-\gamma^2 \pm \sqrt{(\gamma^2-4(1-\Delta)^2)(\gamma^2-4(1+\Delta)^2)}}{2 (1-\Delta^2)}.
    \label{y}
\end{equation}
Substituting (\ref{y}) into (\ref{lambda}), we obtain the exact form of eigenvalues $\lambda$ shown in the main text. These predict EPs at $\gamma_{\rm defect}=2(1\pm\Delta)$ and $\gamma_{\rm defect} = -2(1\pm\Delta)$. Similarly, only the EPs with $\Re(\gamma)>0$ correspond to disspative defect.

However, not all solutions satisfy our localized ansatz $\Re (\kappa)>0$. To be more specific, we separate the real and imaginary parts of $\kappa$ as $\kappa = \kappa_R + i\kappa_I$. For localized states, we require $\kappa_R > 0$. $y = e^{2\kappa} + e^{-2\kappa}$ can be simplified as
\begin{equation}
\begin{aligned}
    & {\rm Re}(y) = 2\cosh (2\kappa_R) \cos (2\kappa_I), \\
    & {\rm Im}(y) = 2\sinh (2\kappa_R) \sin (2\kappa_I) .\label{kappa}
\end{aligned}
\end{equation}
When $\gamma\leq 2(1-\Delta)$, $\Im(y)$ is always $0$, corresponding to two solutions $\kappa_R = 0$ or $\kappa_I = n\pi/2$. Since $|\Re(y_\pm)| \leq 2$, the solutions are $\kappa_R = 0$ and $\kappa_I = \frac 12\arccos (\frac 12 \Re(y_\pm))$. Here, $\kappa$ is purely imaginary. Therefore, the EP located at $\gamma=2(1-\Delta)$ corresponds to the coalescence of the scattering states within one band. 

\begin{figure}
\centering
\includegraphics[width=0.7\linewidth]{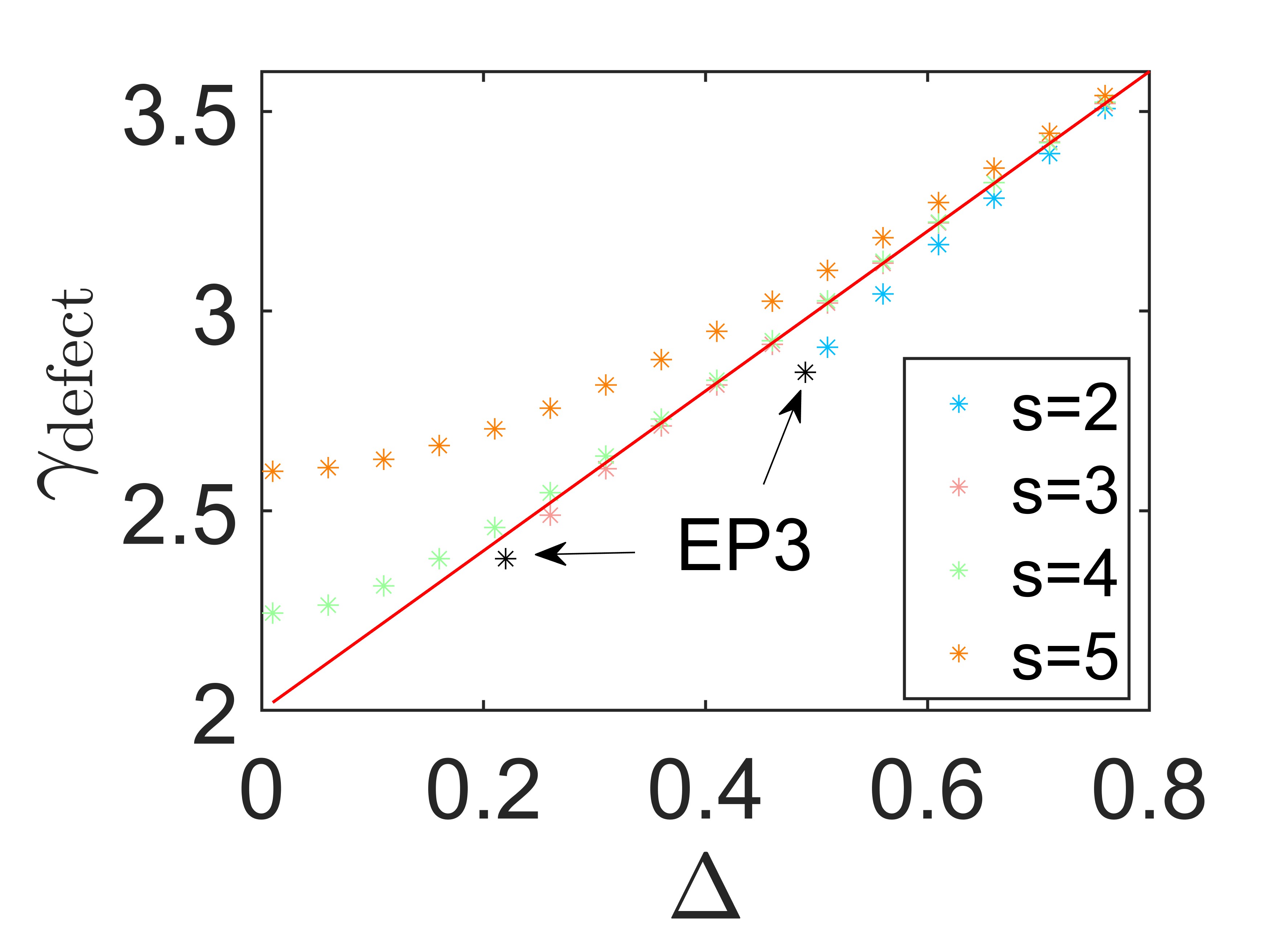}
\caption{\justifying { Comparison between the $\gamma_{\rm defect}$ from (\ref{eq:defectEP}) (lines) and the numerical results (dots) with different defect locations $s$ and a fixed length $2N = 10$. }}
\label{defectEPb}
\end{figure}

When $2(1-\Delta)<\gamma<2(1+\Delta)$,
\begin{equation}
\begin{aligned}
    & \cosh (2\kappa_R) \cos (2\kappa_I) = -\frac {\gamma^2 }{4 (1-\Delta^2)} < -1, \\
    & \sinh (2\kappa_R) \sin (2\kappa_I) = \pm \frac { \sqrt{(\gamma^2-4(1-\Delta)^2)(4(1+\Delta)^2-\gamma^2)}}{4 (1-\Delta^2)},
\end{aligned}
\end{equation}
so that $\kappa_R > 0$. At the EP 
\begin{equation}
    \gamma_{\rm defect} = 2(1+\Delta), \label{eq:defectEP}
\end{equation}
the solutions are $\cosh (2\kappa_R) = r, \kappa_I = \pm\pi/2$. Two mid-gap defect states coalesce.

Finally, when $\gamma>2(1+\Delta)$, $\Im(y)$ is always $0$, and $\Re(|y_\pm|) > 2$. We find the solutions are $\kappa_I = n\pi/2$ and $\kappa_R = \frac 12 \cosh^{-1}(\frac 12 \Re(|y_\pm|))$. Here, two defect states with different localization rates appear, and the energy gap between the them becomes fully imaginary.

Note that, in solving for the defect states, we for simplicity assumed an infinite chain. In a finite chain, the wave functions are truncated near the edge, with approximate form
\begin{equation}
    \psi_i^{\rm cut} \sim 
\left\{
\begin{aligned}
& \psi_i \quad i\in \{1,\cdots,2N\} \\
& 0 \quad\ \rm else
\end{aligned}
\right. ,
\end{equation}
where in the first line $\psi_i$ are the eigenstates we solved for above. The truncation leads to an energy difference
\begin{equation}
    \Delta \lambda \sim \max \{|\psi_0|^2, |\psi_{2N+1}|^2\},
\end{equation}
where
\begin{equation}
\begin{aligned}
    & |\psi_0|^2 = |B|^2 \exp (-2 \kappa_R(s-1)), \\
    & |\psi_{2N+1}|^2 = |A|^2 \exp (-2 \kappa_R(N+1-s)).
\end{aligned}
\end{equation}
The energy difference indicates that the results are more accurate when the defect is far from the boundary, or when the wave functions are more strongly localized. If the defect is closer to the left boundary, the approximation is valid for large $\gamma|a_s|^{-1} \gg 1$, as shown in figure \ref{defectEPb}.

\begin{figure}
\centering
\includegraphics[width=0.7\linewidth]{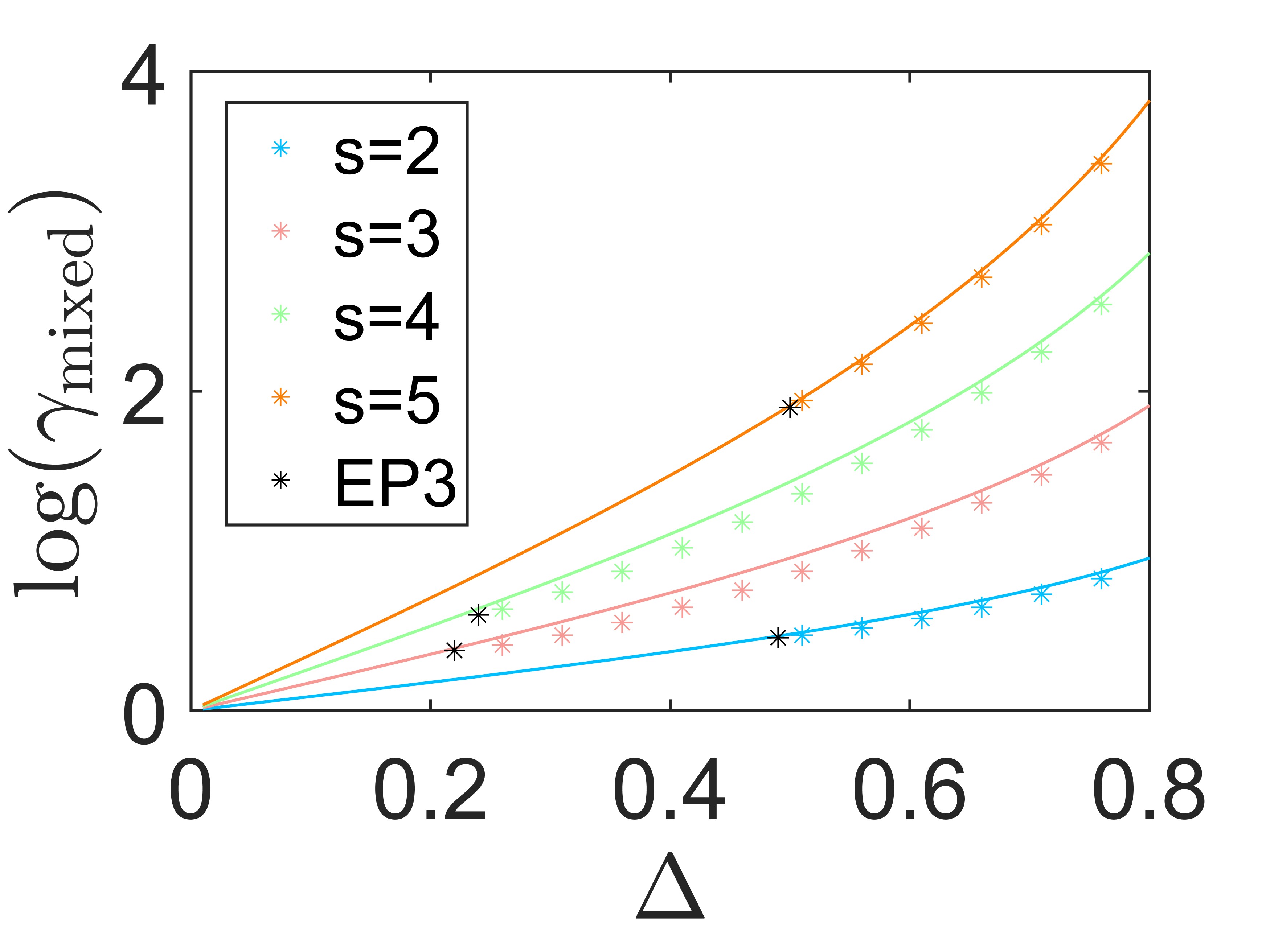}
\caption{\justifying { Comparison between the $\gamma_{\rm mixed}$ from (\ref{eq:mixedEP}) (lines) and the numerical results (dots) with different defect locations $s$ and a fixed length $2N = 10$. }}
\label{mixedEP}
\end{figure}

\subsection{Exceptional points from the hybridization of edge states and defect states}

There is another possibility when a defect state coalesces with an edge state. The existence of such an EP becomes clear by examining the eigenvalues of the four localized states we obtained above. When $\gamma > \max \{\gamma_{\rm edge}, \gamma_{\rm defect}\}$, the four eigenvalues are purely imaginary. When $\gamma \gg \max \{\gamma_{\rm edge}, \gamma_{\rm defect}\}$, these have the approximate form
\begin{equation}
    \lambda_{\rm edge} \approx  -i\gamma|a_s|^2, -i|J|\frac{\gamma_{\rm edge}}{2\gamma} , \quad
    \lambda_{\rm impu} \approx -i\gamma, -i\frac{4\Delta}{\gamma}.
\end{equation}
The gap between one of the edge states and one of the localized states closes at 
\begin{equation}
    \gamma_{\rm mixed} = \sqrt{4\Delta} |a_s|^{-1}. \label{eq:mixedEP}
\end{equation}
After substituting the exact form of $|a_s|$, we arrive at the $\gamma_{\rm mixed}$ in the main text. However, since the eigenenergies of the localized states are largely shifted beyond the approximation region $\gamma|a_s|\ll 1$, the prediction only agrees with the exact results by order of $r^{s-1}$ [see figure \ref{mixedEP}]. 

At the points where $\gamma_{\rm mixed}$ meets with $\gamma_{\rm edge}$ or $\gamma_{\rm defect}$, HOEPs emerge. This also explains the cut off in the plots of the numerically obtained EP values in figures. \ref{edgeEP}-\ref{mixedEP}, where after coalescing at a HOEP the EP2s extend into complex region, out of the plane in these pictures. Since $\gamma_{\rm mixed}$ may coalesce with different EPs when varying the defect location, the values of the HOEP can be solved by two different equations. Specifically, we solve 
\begin{equation}
\begin{aligned}
    & \left\{
    \begin{aligned}
        & \gamma_{\rm mixed} = \gamma_{\rm defect}, & \gamma_{\rm edge} < \gamma_{\rm defect} \\
        & \gamma_{\rm mixed} = \gamma_{\rm edge}, & \gamma_{\rm edge} > \gamma_{\rm defect} 
    \end{aligned}
    \right.
    \Rightarrow \\
    & \left\{
    \begin{aligned}
        & r^{s-1} \sqrt{\frac{1-r^{-2N}}{1-r^{-2}}\cdot 4\Delta} = 2(1+\Delta), & s < \frac{N}{2} + 1  \\
        & r^{s-1} \sqrt{\frac{1-r^{-2N}}{1-r^{-2}}\cdot 4\Delta} =  2(1-\Delta)r^{2s-1-N}, & s > \frac{N}{2} + 1
    \end{aligned}
    \right.
\end{aligned}
\end{equation}
where the choice of different equation is only determined by the relation between defect location $s$ and the sublattice sites $N$. A different case appears at $s = \frac{N}{2} + 1$, where $N$ is even. At this point, we have $\gamma_{\rm edge} = \gamma_{\rm defect}$, which means they have exactly the same value and becomes degenerate. Therefore, the corresponding HOEP is actually an EP4, where all the localized states coalesce.

\nocite{apsrev41Control}
\bibliographystyle{apsrev4-1}
\bibliography{ref}

\end{document}